\documentclass{egpubl}
\usepackage{eg2025}
 
\ConferencePaper        
\CGFStandardLicense

\usepackage[T1]{fontenc}
\usepackage{dfadobe}  

\usepackage{cite}  
\BibtexOrBiblatex
\electronicVersion
\PrintedOrElectronic
\ifpdf \usepackage[pdftex]{graphicx} \pdfcompresslevel=9
\else \usepackage[dvips]{graphicx} \fi

\usepackage{egweblnk} 

\title[Rest Shape Optimization for Sag-Free Discrete Elastic Rods]%
{\vspace{-10mm}Rest Shape Optimization for Sag-Free Discrete Elastic Rods
\vspace{-10mm}
}

\author[T. Takahashi \& C. Batty]
{\parbox{\textwidth}{
\centering 
Tetsuya Takahashi$^{1}$\orcid{0000-0002-9938-5494}
and
Christopher Batty$^{2}$\orcid{0000-0003-3830-7772} 
}
\vspace{-3mm}
\\
{\parbox{\textwidth}{\centering 
$^1$Tencent America, USA\\
$^2$University of Waterloo, Canada
       }
}
\vspace{-12mm}
}

\usepackage{amsmath,bm}
\usepackage{amssymb}
\usepackage{algorithm}
\usepackage{algorithmic}
\usepackage{graphicx}
\usepackage{subcaption}
\usepackage[export]{adjustbox}

\usepackage{pslatex}

\newif\ifdraft
\drafttrue

\ifdraft
\newcommand{\mycolor}[1]{#1}
\newcommand{\note}[1]{\textit{\textcolor{magenta}{#1}}}
\newcommand{\figDir}{low}
\else
\newcommand{\mycolor}[1]{#1}
\newcommand{\note}[1]{}
\newcommand{\figDir}{high}
\fi

\DeclareMathOperator*{\argmin}{arg\,min}

\newcommand{\norm}[1]{\left\lVert#1\right\rVert}
\newcommand{\realNumber}{\mathbb{R}}

\newcommand{\generalizedPos}{\mathbf{q}}
\newcommand{\generalizedVel}{\mathbf{\dot{q}}}
\newcommand{\generalizedForce}{\mathbf{f}}
\newcommand{\generalizedForceExt}{\mathbf{f}_{\mathrm{ext}}}
\newcommand{\generalizedMass}{\mathbf{M}}
\newcommand{\vertexPos}{\mathbf{x}}
\newcommand{\vertexVel}{\mathbf{\dot{x}}}
\newcommand{\edgeAngle}{\boldsymbol{\theta}}
\newcommand{\edgeAngVel}{\boldsymbol{\dot{\theta}}}
\newcommand{\dt}{\Delta t}
\newcommand{\radius}{r}
\newcommand{\length}{\mathbf{l}}
\newcommand{\tangent}{\mathbf{t}}
\newcommand{\curvature}{\boldsymbol{\kappa}}
\newcommand{\twist}{\mathbf{m}}
\newcommand{\gravity}{\mathbf{g}}

\newcommand{\generalizedRestShape}{\mathbf{\bar{s}}}
\newcommand{\generalizedRestShapeMin}{\mathbf{\bar{s}_{\mathrm{min}}}}
\newcommand{\generalizedRestShapeMax}{\mathbf{\bar{s}_{\mathrm{max}}}}
\newcommand{\restLength}{\mathbf{\bar{l}}}
\newcommand{\restCurvature}{\boldsymbol{\bar{\kappa}}}
\newcommand{\restTwist}{\mathbf{\bar{m}}}

\newcommand{\regCoef}{\alpha}
\newcommand{\penaltyCoef}{\beta}
\newcommand{\restCurvatureDelta}{\mathbf{b}}
\newcommand{\restTwistDelta}{\mathbf{d}}

\newcommand{\stretchCoef}{c_{\mathrm{stretch}}}
\newcommand{\bendCoef}{c_{\mathrm{bend}}}
\newcommand{\twistCoef}{c_{\mathrm{twist}}}

\newcommand{\Jacobian}{\mathbf{J}}
\newcommand{\Imat}{\mathbf{I}}
\newcommand{\Hmat}{\mathbf{H}}
\newcommand{\evec}{\mathbf{e}}

\newcommand{\turningAngle}{\phi}
\newcommand{\Amat}{\mathbf{A}}
\newcommand{\Dmat}{\mathbf{D}}

\begin{document}


\teaser{
\vspace{-3mm}
\begin{subfigure}[b]{0.25\linewidth}
\adjincludegraphics[trim={{0.05\width} {0.0\height} {0.05\width} {0.0\height}}, clip, width=\linewidth]
{figures/\figDir/der_0001.png}%
    \vspace{-2mm}
    \caption{Frame 001}
\end{subfigure}%
\begin{subfigure}[b]{0.25\linewidth}
\adjincludegraphics[trim={{0.05\width} {0.0\height} {0.05\width} {0.0\height}}, clip, width=\linewidth]
{figures/\figDir/der_0061.png}%
    \vspace{-2mm}
    \caption{Frame 061}
\end{subfigure}%
\begin{subfigure}[b]{0.25\linewidth}
\adjincludegraphics[trim={{0.05\width} {0.0\height} {0.05\width} {0.0\height}}, clip, width=\linewidth]
{figures/\figDir/der_0115.png}%
    \vspace{-2mm}
    \caption{Frame 115}
\end{subfigure}%
\begin{subfigure}[b]{0.25\linewidth}
\adjincludegraphics[trim={{0.05\width} {0.0\height} {0.05\width} {0.0\height}}, clip, width=\linewidth]
{figures/\figDir/der_0800.png}%
    \vspace{-2mm}
    \caption{Frame 800}
\end{subfigure}%
\vspace{-4mm}
\caption{
Our rest shape optimizer enables sag-free simulation of discrete-elastic-rod-based hair strands. In this scene where the motions of the root vertices are driven by a rotating sphere, our method preserves the carefully designed hair style without stiffening the strands, and achieves natural hair dynamics under gravity. We use 1.9K strands, and each strand is discretized with 100 vertices. Our rest shape optimization took 11.2 s (with 7.6 Gauss-Newton iterations on average for each strand) while forward simulation (without collision handling) took 3.3 s per frame.
}
\label{fig:teaser}
}

\maketitle

\vspace{-2mm}
\begin{abstract}
We propose a new rest shape optimization framework to achieve sag-free simulations of discrete elastic rods. To optimize rest shape parameters, we formulate a minimization problem based on the kinetic energy with a regularizer while imposing box constraints on these parameters to ensure the system's stability. Our method solves the resulting constrained minimization problem via the Gauss-Newton algorithm augmented with penalty methods. We demonstrate that the optimized rest shape parameters enable discrete elastic rods to achieve static equilibrium for a wide range of strand geometries and material parameters.


\vspace{-2mm}
\end{abstract}

\vspace{-4mm}
\section{Introduction}
\vspace{-2mm}
From human hairs to cables for electronic devices, elastic strands can be found everywhere in our daily lives, and modeling such thread-like materials is of critical importance in various domains, including computer graphics, mechanical engineering, and medical applications. The Discrete Elastic Rods (DER) formulation \cite{Bergou2008,Bergou:2010:DVT:1778765.1778853} is one of the most popular approaches to accurately and efficiently simulate such elastic strands and has demonstrated its effectiveness in a wide range of applications involving both forward simulations \cite{Chentanez2009needle,DiVerdi2010,AUDOLY201318,Kaufman2014,Schweickart2017,Fei2019,Lesser2022,Daviet2023hair} and inverse problems \cite{Perez2015rod,Zehnder2016curve,Perez2017surfaces,Schumacher2018sheet,Malomo2018flexmaps,Xu2018wire,Panetta2019,Neveu2022curve,Ren2022umbrella,chen2024differentiablediscreteelasticrods}.

When modeling and designing elastic strands, a well-known problem is that such strands will immediately sag due to gravity when the simulation starts \cite{Twigg:2011:OSS:2019406.2019437,Hsu2023sag}, which ruins the specific strand shapes (e.g., hairstyles) that were either carefully designed by users, technical artists, and designers \cite{Iben2019,Liu2022gravity} or captured from real human hair or wigs \cite{Luo2013hair}. This issue occurs because the simplest way to model the strands is to focus solely on the geometry, neglecting the role of forces or dynamics; that is, the strand's initial shape is treated as its rest shape \cite{Kim2022deformables}. While one can largely mitigate this sagging effect by significantly (and artificially) increasing the material stiffness, this approach makes the elastic strands excessively stiff and sacrifices their natural dynamic response, thus negatively impacting the visual result.

To address this sagging problem in the DER framework, we propose a rest shape optimization approach that optimizes the rest shape parameters (rest length, rest curvature, and rest twist) of the input strands such that the input shape is preserved at static equilibrium under simulation (i.e., the strand shape deformed due to simulation coincides with the input shape). To find optimal rest shape parameters, we formulate a minimization problem based on the kinetic energy. Given the greater number of degrees of freedom (DOFs) for the rest shape parameters compared to the DOFs of DER strands, we incorporate a regularizer to avoid creating underdetermined systems. In addition, we impose box constraints on the rest shape parameters to prevent significant changes from their original values, ensuring the stability of the system. We solve the resulting constrained nonlinear minimization problem via the Gauss-Newton (GN) algorithm \mycolor{(and thus without the Hessian of the DER objectives)} augmented with penalty methods and line search. Fig. \ref{fig:teaser} demonstrates the efficacy of our rest shape optimization.

\vspace{-4mm}
\section{Related Work}
\vspace{-2mm}

\subsection{Elastic Rod Simulation}
\vspace{-2mm}
Due to the one-dimensional structure of thin elastic strands, it is typically difficult to accurately simulate these materials with general finite element methods (FEM) designed for volumetric structures \cite{Kim2022deformables}, and specialized approaches have been investigated. Given that common strand-like materials exhibit higher stiffness in stretching modes compared to bending and twisting, early work assumed inextensibility via e.g., constraints based on Lagrange multipliers \cite{Goldenthal2007cloth} with linear-time factorization and triangular solves \cite{Baraff1996linear}, a tridiagonal matrix formulation \cite{Han2013tridiagonal}, and geometric stiffness \cite{Tournier2015,Andrews2017stiffness}, reduced multi-body solvers \cite{Hadap2006strands,Featherstone2016}, and position-based dynamics (PBD) \cite{Muller:2007:PBD:1234415.1234575,Bender2017survey}. To support the characteristic dynamics of elastic strands, the traditional mass-spring framework \cite{Liu:2013:FSM:2508363.2508406} was extended with altitude springs \cite{Selle2008mass}, bending springs \cite{Iben2013hair}, torsion energy \cite{Shi2023curls}, and ghost rest shapes \cite{Herrera2024}. To simulate one-dimensional strands along with other deformables in a unified way, supporting torsional effects is critical, and therefore discretization elements that can capture twists (e.g., oriented particles \cite{Muller2011oriented} and elastons \cite{Martin2010unified}) have also been presented. For highly constrained strand-like systems, further specialized approaches have been proposed with, e.g., Eulerian nodes \cite{Sueda2011strands,Sanchez2020rods}, long range constraints \cite{Muller2017rigid}, and unilateral distance constraints \cite{Muller2018cable}.

Since the introduction of the Cosserat theory into graphics \cite{Pai2002cosserat}, it has been adapted to dynamical systems of strands \cite{Spillmann2007rod} and network structures \cite{Spillmann2009nets}, combined with PBD using ghost particles \cite{Umetani2014rod}, and further extended to support edge rotations via quaternions in the PBD framework \cite{Kugelstadt2016rod}. The work of Kugelstadt and Sch\"{o}mer~\cite{Kugelstadt2016rod} has been extended to remove the ambiguity in the Darboux vector \cite{Hsu2023sag}, to handle stiff problems via compliant constraints while utilizing the tree structures \cite{Deul2018rod}, to preserve the volume of elastic strands \cite{Angles2019rods}, and to achieve consistent dynamics \cite{Soler2018rod} using projective dynamics \cite{Bouaziz:2014:PDF:2601097.2601116}. Elastic rod simulation based on the Cosserat theory \cite{Spillmann2007rod} has also been extended to support rh-adaptivity \cite{Wen2020rod} and a compact representation of the DOFs \cite{Zhao2022rod}. A curvature-based reduced representation has also been presented by Bertails et al.~\cite{Bertails2006hair} and extended to generate smoother dynamics via clothoids \cite{Casati2013clothoids}.

While various approaches have been presented for simulating elastic strands, DER \cite{Bergou2008,Bergou:2010:DVT:1778765.1778853}, which discretizes an elastic strand with vertex positions and edge angles in maximal coordinates, is one of the most popular approaches due to its simplicity, efficiency, and flexibility, as employed in a wide range of applications. As such, we use DER as our elastic strand simulator.


\vspace{-4mm}
\subsection{Sag-Free Simulation}
\vspace{-2mm}
Sagging of elastic materials at the onset of simulation has been a notorious and challenging problem; resolving it requires ensuring static equilibrium during the modeling, design, and fabrication of such materials. Thus, various methods have been proposed to achieve sag-free simulation. We classify these methods into the following three categories: nonlinear force solve, constrained minimization,  and global-local initialization approaches.


\vspace{-3mm}
\subsubsection{Nonlinear Force Solve}
\vspace{-2mm}
Solving Newton's second law of motion for (typically nonlinear) forces and zero acceleration is equivalent to achieving zero net force and thus static equilibrium. Hadap \cite{Hadap2006strands} proposed using \mycolor{inverse dynamics \cite{Featherstone2016} to solve the nonlinear equation on the forces and thus enable sag-free simulation. While inverse dynamics works for reduced multibody systems, in general, it is not clear how to apply it for general simulation systems represented in maximal coordinates (e.g., FEM-based deformables and DER)}. 

The nonlinear equations on the forces can be solved, e.g., with Newton-type optimizers \cite{Wang:2015:DCM:2809654.2766911,Mukherjee2018shape}, although Iben \cite{Iben2019} aimed to cancel gravity using spring forces via local analytical solutions within the mass-spring systems \cite{Iben2013hair}. However, in practice, it is necessary to consider a global system to fully cancel gravity. A similar technique was also employed for sag-free mass-spring systems by introducing additional artificial springs (which can potentially introduce unnatural forces) \cite{Herrera2024}.

Chen et al. \cite{Chen2014asymptotic} employed an Asymptotic Numerical Method (ANM), which is a homotopy approach to solving nonlinear equations, for inverse elastic shape design with incompressible neo-Hookean materials. They demonstrated orders of magnitude faster convergence than one of the Newton-type optimizers, Levenberg–Marquardt algorithm (LMA) \cite{NoceWrig06}. Later, Jia \cite{Jia2021SANM} extended ANM to support more general material types and also demonstrated its efficiency for forward simulation compared to optimization-based integrators \cite{Martin:2011:EEM:2010324.1964967,Gast2015} combined with Newton's method \cite{NoceWrig06} (although ANM can yield solutions that are local maxima, unlike those generated by the optimization-based integrators). ANM was also employed to investigate the equilibrium of elastic rods \cite{LAZARUS20131712}. While ANM can be more efficient than the Newton-type optimizers \cite{NoceWrig06}, it is not clear how to support box constraints, which are indispensable to constrain rest shape parameters within physically valid ranges and avoid introducing stability issues.

\mycolor{
While canceling gravity forces in DER has been previously attempted by solving force equations or equivalent minimization \cite{Aubry2015,Lesser2022}, their approach lacked algorithmic and formulation specifics, such as exact objectives and DER force Jacobians with respect to the rest shape parameters. Additionally, their approach did not support box constraints which are critical to avoid significant rest shape changes and thus ensure stable simulations.
}


\vspace{-3mm}
\subsubsection{Constrained Minimization}
\vspace{-2mm}
Derouet-Jourdan et al.~\cite{Derouet-Jourdan2010} presented a method for achieving static equilibrium of elastic strands, simulated with the curvature-based discretization \cite{Bertails2006hair}, in 2D. They formulated an objective consisting of curvatures and material parameters and then optimized these parameters via linear solves on the curvature and stiffening/lightening of materials. Derouet-Jourdan et al.~\cite{Derouet-Jourdan:2013:IDH:2508363.2508398} also achieved static equilibrium of curvature-based hair strands \cite{Bertails2006hair} in 3D under frictional contacts. Along the same lines, inverse approaches for curvature-based elastic strands have been further investigated \cite{bertailsdescoubes:hal-01827887,Charrondiere2020,Hafner2021curves,Hafner2023rod}.

Twigg and Ka\v{c}i\'{c}-Alesi\'{c} proposed optimizing rest shape parameters via force-norm minimization to achieve static equilibrium of mass-spring systems \cite{Twigg:2011:OSS:2019406.2019437}. Our work is most closely related to theirs, so we clarify the differences in detail in Sec.\ \ref{sec:differences_from_twigg}. Skouras et al.~\cite{Skouras2012design} presented an optimization approach that treats the nonlinear force equations as hard constraints and solves the optimization via the augmented Lagrangian method \cite{NoceWrig06}. Similar optimization problems under hard constraints have been proposed and efficiently addressed using the adjoint method \cite{Perez2015rod,Zehnder2016curve,Perez2017surfaces,Ly2018shell,Malomo2018flexmaps,Panetta2019}. While these adjoint-based approaches are efficient \mycolor{and can ensure local minima (not maxima)}, their implementation tends to be significantly more complicated because the nonlinear force constraints need to be differentiated, thus involving the Hessian of the objectives of elastic materials. Besides complexity, prior work has highlighted additional challenges introduced by this Hessian, related to correctness, accuracy, robustness, and efficiency \cite{Panetta2019,Shi2023curls}. As such, similar to the work of \cite{Twigg:2011:OSS:2019406.2019437}, we formulate our method without explicitly encoding the Hessian into the rest shape optimization. To avoid these complications for the inversions, Choi et al.~\cite{Choi2024} proposed using the finite difference method to approximate the gradient with respect to the curvatures while Liu \cite{Liu2022gravity} treated a simulator as a black-box using search directions approximated with deformation gradients, although such schemes are known to be rather inefficient.


\vspace{-3mm}
\subsubsection{Global-Local Initialization}
\label{sec:global_local}
\vspace{-2mm}
Hsu et al.~\cite{Hsu2022sag} proposed a global-local, two-stage initialization method that first solves a global linear system to find optimal forces and then locally adjusts parameters on the rest shape, material stiffness, and internal state such that these parameters become consistent with the computed forces. Compared to the work of Twigg and Ka\v{c}i\'{c}-Alesi\'{c}~\cite{Twigg:2011:OSS:2019406.2019437} and our own, which need to solve a global nonlinear system iteratively, the two-stage initialization \cite{Hsu2022sag} requires solving a global linear system only once which reduces the computational cost significantly. This two-stage initialization was also extended to support hair simulation \cite{Hsu2023sag} with the elastic rod model of \cite{Kugelstadt2016rod}. In practice, while the two-stage initialization is efficient, forces computed via the single global linear solve are not necessarily feasible with only the modified rest shape parameters of local elements. As such, one would typically need to compromise with non-zero forces failing to achieve static equilibrium or locally modified material stiffness parameters deviating from the user-desired strand dynamics, as discussed in \cite{Hsu2022sag,Hsu2023sag}. 

In addition, it is nontrivial to apply the global-local initialization approach \cite{Hsu2022sag,Hsu2023sag} to DER because its bending and twisting formulations do not satisfy the requirements of the global-local initialization. First, the bending and twisting formulations are tightly coupled in DER, and thus the local (minimal) stencil for bending does not necessarily preserve the angular momentum by itself (we confirmed this with the optimal forces generated by our method). This fact makes it difficult to define \emph{force elements}, which are used to form the global linear system. Furthermore, assuming the use of the local stencil as a force element, the bending/twisting couples the rest length parameters with neighboring ones, transforming the (ostensibly) local steps into a global problem. \mycolor{As such, it is necessary to modify the global-local initialization to apply it to DER although such an approach would fail to achieve static equilibrium (see Sec. \ref{sec:hair_like}).}


\vspace{-3mm}
\subsubsection{Differences from \cite{Twigg:2011:OSS:2019406.2019437}}
\label{sec:differences_from_twigg}
\vspace{-2mm}
Twigg and Ka\v{c}i\'{c}-Alesi\'{c}~\cite{Twigg:2011:OSS:2019406.2019437} proposed a rest shape optimization framework for sag-free mass-spring systems. Their method formulates an objective based on the $L^2$ norm of forces with a related regularizer, and optimizes the rest length and rest angle under box constraints for the mass-spring networks. While this work has a goal similar to ours, there are two key differences. First, instead of mass-spring systems, we use DER (which has been validated through physical experiments to generate dynamics in accordance with theory \cite{Bergou2008,Bergou:2010:DVT:1778765.1778853,jawed2018primer,Romero2021validation}) and therefore we optimize the rest shape parameters of the DER objectives: rest length, rest curvature, and rest twist. In addition, we also impose corresponding box constraints (based on our analysis and numerical experiments) on the rest shape parameters to prevent significant changes in their values and thus avoid introducing stability issues. Second, instead of the $L^2$ norm of forces, we formulate an objective based on kinetic energy. While these are conceptually similar, our kinetic-energy-based objective can correctly reflect the mass/inertia of materials and consistently evaluate the energy contributions of each component in the system. Moreover, our kinetic-energy-based formulation has much better numerical conditioning than the force-norm-based one \mycolor{due to the proper consideration of the energy contributions} (see Sec. \ref{sec:kinetic_energy_vs_force_norm}), making it possible to use double-precision floating-point when solving the optimization problem. By contrast, the force-norm-based objective can be numerically ill-conditioned such that, \mycolor{to achieve sufficiently small error} during the optimization, extended-precision \mycolor{would be} required (such as the quad-precision adopted in \cite{Twigg:2011:OSS:2019406.2019437}). These features are not standard nor officially supported in C++20, for example, and significantly slow down the computations.


\vspace{-4mm}
\section{Discrete Elastic Rods Preliminaries}
\label{sec:DER}
\vspace{-2mm}
In preparation for the rest shape optimization approach we develop in Sec.~\ref{sec:rest_shape_optimization}, we briefly review the energy objectives in the DER formulation and their gradients, which play key roles in forward simulation \cite{Bergou2008,Bergou:2010:DVT:1778765.1778853}. Here, as our focus is on the static equilibrium case, we do not consider \mycolor{contacts,} plastic deformations, or tearing, and cover details only directly relevant to our rest shape optimizer. We refer readers to the book on DER \cite{jawed2018primer} for the underlying principles including space/time-parallel transport  and to the previous work for derivation and details of the gradient and Hessian \cite{Panetta2019,Fei2019}. In addition, because we consider scenarios without any contacts among elastic rods and other objects, each rod can independently be processed in both forward simulation and rest shape optimization. As such, we describe formulations for a single elastic rod below. 

We consider an elastic rod discretized with $N$ vertices (which are sufficiently well distributed without forming acute angles) whose stacked positions are denoted by $\vertexPos = [\vertexPos_0^T, \ldots, \vertexPos_{N-1}^T]^T \in \realNumber^{3N}$ (each vertex position is three-dimensional) and $(N-1)$ connecting edges whose stacked edge angles are $\edgeAngle = [\edgeAngle_0, \ldots, \edgeAngle_{N-2}]^T \in \realNumber^{N-1}$ (each edge angle is one-dimensional). The generalized positions of the rod can be defined by interleaving the vertex positions and edge angles as $\generalizedPos = [\vertexPos_0^T, \edgeAngle_0, \ldots, \edgeAngle_{N-2}, \vertexPos_{N-1}^T]^T \in \realNumber^{4N-1}$. This specific variable arrangement of $\generalizedPos$ leads to a banded Hessian for the DER objectives with  time-parallel transport (which also accommodates trivial parallel computations \cite{Kaldor2010}) and thus enables efficient implicit integration \cite{Bergou:2010:DVT:1778765.1778853}, via Cholesky/LDLT-based direct linear solves \cite{Huang2023hair}. Along with the positional variables, we also define stacked velocities for the vertices as $\vertexVel = [\vertexVel_0^T, \ldots, \vertexVel_{N-1}^T]^T \in \realNumber^{3N}$, stacked angular velocities for the edges $\edgeAngVel = [\edgeAngVel_0, \ldots, \edgeAngVel_{N-2}]^T \in \realNumber^{N-1}$, and generalized velocities $\generalizedVel = [\vertexVel_0^T, \edgeAngVel_0, \ldots, \edgeAngVel_{N-2}, \vertexVel_{N-1}^T]^T \in \realNumber^{4N-1}$. For simplicity, we assume perfectly circular cross-sections with a constant radius, constant density, and constant material stiffness over the strand. In addition, we initialize reference frames on each edge via space-parallel transport propagated from a randomly chosen reference frame on the first edge (i.e., any reference frame on the first edge, as long as it forms the $SO(3)$ basis with the unit tangent vector, gives identical results for the dynamics) and then compute material frames based on the reference frames and edge angles $\edgeAngle$ \cite{Bergou2008,Bergou:2010:DVT:1778765.1778853,jawed2018primer}. While, in the previous work \cite{Bergou2008,Bergou:2010:DVT:1778765.1778853,jawed2018primer}, subscript and superscript indices are used for the variables defined on vertices and edges, respectively, we use only subscript indices to avoid confusion with common superscripts (e.g., exponentiation, inverse, transpose, and time index), and because we can identify whether variables are defined on a vertex or an edge from their definition.

While the dimensionality of $\generalizedPos$ is $(4N-1)$ in 3D, we focus on an elastic rod with its root end minimally clamped as this is one of the simplest and most common settings, e.g., for hairs attached to a head (although it would be possible to support both ends clamped \cite{Hafner2023rod}). To this end, we fix $\vertexPos_0, \edgeAngle_0$, and $\vertexPos_1$ or update them in a prescribed way (below, we use the term ``fix'' regardless of whether these variables are fixed in place or moved in a prescribed way) so that the elastic rod has an anchor enabling stretching/bending/twisting to work meaningfully, i.e., these variables work as a Dirichlet boundary condition to eliminate DOFs for pure rigid motions. This treatment corresponds to eliminating the DOFs for $\vertexPos_0, \edgeAngle_0$, and $\vertexPos_1$, and thus under our minimal root-end clamping the resulting number of free DOFs is $(4N-8) = (4N-1) - (3 + 1 + 3)$. 

A single forward simulation step of DER can be written as an energy minimization problem \cite{Martin:2011:EEM:2010324.1964967,Gast2015}:
\begin{align}
\generalizedPos = \argmin_{\generalizedPos} E_{\mathrm{DER}}(\generalizedPos),
\label{eq:forward_sim_optimization}
\end{align}
where the objective of DER, $E_{\mathrm{DER}}(\generalizedPos)$ consists of the objective terms for inertia $E_{\mathrm{inertia}}(\generalizedPos)$, stretching $E_{\mathrm{stretch}}(\vertexPos)$, bending $E_{\mathrm{bend}}(\generalizedPos)$, and twisting $E_{\mathrm{twist}}(\generalizedPos)$, and can be defined as
\begin{align}
E_{\mathrm{DER}}(\generalizedPos) = 
E_{\mathrm{inertia}}(\generalizedPos) +
E_{\mathrm{stretch}}(\vertexPos) +
E_{\mathrm{bend}}(\generalizedPos) +
E_{\mathrm{twist}}(\generalizedPos).
\end{align}
Here, we define $E_{\mathrm{stretch}}(\vertexPos)$ with $\vertexPos$ alone because stretching is independent of $\edgeAngle$. While velocity-based damping can also be incorporated into the optimization \eqref{eq:forward_sim_optimization} \cite{George2018}, it is irrelevant to our rest shape optimization since we focus on the static equilibrium case \mycolor{with $\generalizedVel=0$}. We can obtain the generalized positions for the next time step by solving the optimization problem \eqref{eq:forward_sim_optimization}, \mycolor{e.g., with a gradient-based optimizer (such as gradient descent or Newton's method \cite{NoceWrig06})} using the gradient of the DER objective with respect to $\generalizedPos$, $\nabla_{\generalizedPos}E_{\mathrm{DER}}(\generalizedPos)$, while handling the fixed variables as Dirichlet boundary conditions. We note that the gradient with respect to the fixed variables is not defined (or, in practice, it can be assumed 0 for convenience). In our framework, we use a single Newton iteration per simulation step for efficiency \mycolor{\cite{Baraff1998,Kim2022deformables,Huang2023hair}} (although this approach sacrifices stability with larger parameter values \cite{Shi2023curls}). \mycolor{If one prefers avoiding the Hessian of the DER objectives (to make the entire framework Hessian-free), it would be possible to use Alternating Direction Method of Multipliers (ADMM) \cite{Narain2016admm} to solve \eqref{eq:forward_sim_optimization}.} In the following sections, we define each objective term and its gradient. 


\vspace{-4mm}
\subsection{Inertia}
\label{sec:DER_inertia}
\vspace{-2mm}
We define the objective for the inertia as
\begin{align}
E_{\mathrm{inertia}}(\generalizedPos) =
\frac{1}{2\dt^2}\norm{\generalizedPos - \generalizedPos^*}^2_{\generalizedMass},
\label{eq:inertia_objective}
\end{align}
where $\generalizedPos^* (= 
\generalizedPos^t + 
\dt \generalizedVel^t + 
\dt^2 \generalizedMass^{-1} \generalizedForceExt)$ denotes the predicted generalized positions, $\generalizedPos^t$ and $\generalizedVel^t$ denote the generalized positions and velocities at time $t$, respectively, $\dt \ (> 0)$ time step size, $\generalizedMass \in \realNumber^{(4N-1) \times (4N-1)}$ a diagonal, generalized mass matrix \cite{Bergou:2010:DVT:1778765.1778853,jawed2018primer}, and $\generalizedForceExt \in \realNumber^{4N-1}$ a generalized external force vector. While sag-free simulation typically assumes gravity as the sole external force, we can nevertheless easily support other constant external forces, such as wind forces or load on specific vertices \cite{Perez2015rod} or edges. 

The diagonal mass matrix $\generalizedMass$ is defined as $\mathrm{diag}(\generalizedMass) = [m_0, m_0, m_0, I_0, \ldots, I_{N-2}, m_{N-1}, m_{N-1}, m_{N-1}]^T$, where $m_i$ denotes the mass for vertex $i$, and $I_i$ the inertia for edge $i$. Specifically, $m_0 = m_1 = \infty$ (as the first two root-end vertices are fixed), and other $m_i$ can be computed by
\begin{align}
m_i = \rho \pi \radius^2 \left(\frac{\restLength_{i-1} + \restLength_{i}}{2}\right),
\label{eq:mass}
\end{align}
where $\rho$ and $\radius$ denote the mass density and the radius of the strand, respectively, $\restLength_i \ (> 0)$ denotes the rest length for edge $i$ (we define $\restLength = \left[\restLength_0, \ldots, \restLength_{N-2}\right]^T \in \realNumber^{N-1}$), and we specify $\restLength_{-1} = \restLength_{N-1} = 0$ for convenience. Then, $I_0 = \infty$, and other $I_i$ are computed by
\begin{align}
I_i = \frac{1}{2}\rho \pi \radius^4 \restLength_i.
\label{eq:inertia}
\end{align}

The gradient of $E_{\mathrm{inertia}}(\generalizedPos)$ with respect to $\generalizedPos$ (i.e., $\nabla_{\generalizedPos}E_{\mathrm{inertia}}(\generalizedPos) \in \realNumber^{4N-1}$) is given by
\begin{align}
\nabla_{\generalizedPos}E_{\mathrm{inertia}}(\generalizedPos) = 
\frac{\generalizedMass}{\dt^2}(\generalizedPos - \generalizedPos^*).
\label{eq:inertia_gradient}
\end{align}


\vspace{-4mm}
\subsection{Stretching}
\label{sec:DER_stretch}
\vspace{-2mm}
The stretching energy is defined over each edge using its associated two vertices (i.e., both ends of the edge). Thus, the objective for stretching can be decomposed and defined as
\begin{align}
E_{\mathrm{stretch}}(\vertexPos) = 
\sum_{i = 1}^{N-2} E_{\mathrm{stretch}, i}(\vertexPos_{i}, \vertexPos_{i+1}),
\end{align}
and the stretching energy defined for edge $i$ with the associated vertices $i$ and $i+1$ is given as 
\begin{align}
E_{\mathrm{stretch}, i}(\vertexPos_{i}, \vertexPos_{i+1}) = 
\frac{1}{2}\left(\frac{\stretchCoef \pi \radius^2}{\restLength_i}\right)(\length_i - \restLength_i)^2,
\end{align}
where $\stretchCoef$ denotes the stiffness coefficient for stretching, and $\length_i \ (= \norm{\vertexPos_{i+1} - \vertexPos_{i}}_2)$ denotes the length of edge $i$. (Unlike the work of \cite{Bergou:2010:DVT:1778765.1778853}, which uses Young's modulus and shear modulus, we use different coefficients for each energy to individually control their strengths.) Given that the vertices $\vertexPos_{0}$ and $\vertexPos_{1}$ are fixed, there is no stretching energy contribution $E_{\mathrm{stretch}, 0}(\vertexPos_{0}, \vertexPos_{1})$ from edge $0$ (the first edge at the root end). 

The gradient can be computed as
\begin{align}
\nabla_{\vertexPos_{i+1}} 
E_{\mathrm{stretch}, i}(\vertexPos_{i}, \vertexPos_{i+1}) 
&= -
\nabla_{\vertexPos_{i}} 
E_{\mathrm{stretch}, i}(\vertexPos_{i}, \vertexPos_{i+1})
\\
&= \stretchCoef \pi \radius^2 (\length_i \restLength_i^{-1} - 1) \tangent_i,
\label{eq:stretch_grad}
\end{align}
where $\tangent_i \ (= \frac{\vertexPos_{i+1} - \vertexPos_{i}}{\length_i})$ denotes the unit tangent vector for edge $i$.


\vspace{-4mm}
\subsection{Bending}
\label{sec:DER_bend}
\vspace{-2mm}
The bending energy is defined on an inner vertex (excluding the first and last vertices) together with its two neighboring edge angles and vertices, i.e., with $(\vertexPos_{i-1}, \edgeAngle_{i-1}, \vertexPos_{i}, \edgeAngle_{i}, \vertexPos_{i+1}) \in \realNumber^{11}$. Thus, the objective for bending can be decomposed and defined as
\begin{align}
E_{\mathrm{bend}}(\generalizedPos) &= 
\sum_{i = 1}^{N-2} E_{\mathrm{bend}, i}
(\vertexPos_{i-1}, \edgeAngle_{i-1}, \vertexPos_{i}, \edgeAngle_{i}, \vertexPos_{i+1}).
\end{align}
We define the bending energy on vertex $i$ with the four-dimensional curvature $\curvature_i \in \realNumber^{4}$ (\mycolor{$\curvature_{i} = [\curvature_{i, 0}, \curvature_{i, 1}, \curvature_{i, 2}, \curvature_{i, 3}]^T$, where two 2D curvatures $(\curvature_{i, 0}$, $\curvature_{i, 1}$) and ($\curvature_{i, 2}$, $\curvature_{i, 3}$) denote the curvature binormals evaluated on the material frames of edges $i-1$ and $i$, respectively, and} $\curvature = [\curvature_{1}^T, \ldots, \curvature_{N-2}^T]^T \in \realNumber^{4N-8}$) and its corresponding rest curvature $\restCurvature_i \in \realNumber^{4}$ in 3D \cite{Bergou2008,Fei2019} as
\begin{align}
E_{\mathrm{bend}, i}
(\vertexPos_{i-1}, \edgeAngle_{i-1}, \vertexPos_{i}, \edgeAngle_{i}, \vertexPos_{i+1}) = 
\frac{1}{2}
\left(\frac{\bendCoef \pi \radius^4}{4 (\restLength_{i-1} + \restLength_{i})}\right)
\norm{\curvature_i - \restCurvature_i}_2^2,
\label{eq:bending_i}
\end{align}
where $\bendCoef$ denotes the stiffness coefficient for bending energy. Here, we do not use the bending model with the two-dimensional curvature (and rest curvature) via the averaged material frames from \cite{Bergou:2010:DVT:1778765.1778853} because their model would fail to correctly evaluate the bending of strands \mycolor{when they are bent or twisted due to the non-unit and non-orthogonal, averaged material frames} \cite{Gornowicz2015,Panetta2019}. In addition, the bending model \eqref{eq:bending_i} \cite{Bergou2008,Fei2019} is slightly simpler than the one presented by \cite{Panetta2019} (which modifies the rest length term) and avoids introducing additional complexity when differentiating \eqref{eq:bending_i} with respect to the rest length (see Sec. \ref{sec:rest_shape_optimization_bend}).

The 11-dimensional gradient of \eqref{eq:bending_i} can be computed by
\begin{align}
\nabla E_{\mathrm{bend}, i}
(\vertexPos_{i-1}, \edgeAngle_{i-1}, \vertexPos_{i}, \edgeAngle_{i}, \vertexPos_{i+1}) 
= 
\left(\frac{\bendCoef \pi \radius^4}{4 (\restLength_{i-1} + \restLength_{i})}\right)
\Jacobian_{\mathrm{curv}, i}^T (\curvature_i - \restCurvature_i),
\label{eq:bending_i_grad}
\end{align}
where $\Jacobian_{\mathrm{curv}, i} \in \realNumber^{4 \times 11}$ denotes the Jacobian of $\curvature_i$ with respect to the variables $(\vertexPos_{i-1}, \edgeAngle_{i-1}, \vertexPos_{i}, \edgeAngle_{i}, \vertexPos_{i+1})$, and $\Jacobian_{\mathrm{curv}, i}^T (\curvature_i - \restCurvature_i) \in \realNumber^{11}$ in \eqref{eq:bending_i_grad} can be rewritten with the gradient of $\curvature_i$'s $j$th entry, $\curvature_{i, j}$ (i.e., $\nabla \curvature_{i, j} \in \realNumber^{11}$) as
\begin{align}
\Jacobian_{\mathrm{curv}, i}^T (\curvature_i - \restCurvature_i) =
\sum_{j=0}^{3}(\curvature_{i, j} - \restCurvature_{i, j}) 
\nabla \curvature_{i, j}.
\end{align}


\vspace{-4mm}
\subsection{Twisting}
\label{sec:DER_twist}
\vspace{-2mm}
Similar to the bending case, the objective for twisting can be defined with $(\vertexPos_{i-1}, \edgeAngle_{i-1}, \vertexPos_{i}, \edgeAngle_{i}, \vertexPos_{i+1}) \in \realNumber^{11}$ as
\begin{align}
E_{\mathrm{twist}}(\generalizedPos) &= 
\sum_{i = 1}^{N-2} E_{\mathrm{twist}, i}
(\vertexPos_{i-1}, \edgeAngle_{i-1}, \vertexPos_{i}, \edgeAngle_{i}, \vertexPos_{i+1}),
\end{align}
where the twisting energy defined for vertex $i$ is given with twist $\twist_i$ ($\twist = [\twist_{1}, \ldots, \twist_{N-2}]^T \in \realNumber^{N-2}$), which takes the reference twist into account \cite{Bergou:2010:DVT:1778765.1778853,Kaldor2010,jawed2018primer,Panetta2019}, and its corresponding rest twist $\restTwist_i$ by
\begin{align}
E_{\mathrm{twist}, i}
(\vertexPos_{i-1}, \edgeAngle_{i-1}, \vertexPos_{i}, \edgeAngle_{i}, \vertexPos_{i+1}) = 
\frac{1}{2}
\left(\frac{\twistCoef \pi \radius^4}{(\restLength_{i-1} + \restLength_{i})}\right)
(\twist_i - \restTwist_i)^2,
\label{eq:twist_i}
\end{align}
where $\twistCoef$ denotes the stiffness coefficient for the twisting energy.

The 11-dimensional gradient of \eqref{eq:twist_i} can be computed as
\begin{align}
\nabla E_{\mathrm{twist}, i}
(\vertexPos_{i-1}, \edgeAngle_{i-1}, \vertexPos_{i}, \edgeAngle_{i}, \vertexPos_{i+1}) = 
\left(\frac{\twistCoef \pi \radius^4}{(\restLength_{i-1} + \restLength_{i})}\right)
(\twist_i - \restTwist_i) \nabla \twist_i.
\label{eq:twist_i_grad}
\end{align}


\vspace{-4mm}
\section{Rest Shape Optimization}
\label{sec:rest_shape_optimization}
\vspace{-2mm}
We define the rest shape parameter $\generalizedRestShape = [\restLength_1, \ldots, \restLength_{N-2}, \restCurvature^T, \restTwist^T]^T \in \realNumber^{6N-12}$ since $(6N - 12) = (N - 2) + 4(N - 2) + (N - 2)$. We aim to optimize $\generalizedRestShape$ based on minimization of the kinetic energy with a regularizer subject to box constraints. Specifically, we formulate the rest shape optimization as
\begin{align}
\generalizedRestShape = 
\argmin_{
\generalizedRestShapeMin \leq 
\generalizedRestShape \leq 
\generalizedRestShapeMax} \tilde{F}(\generalizedRestShape),
\quad
\tilde{F}(\generalizedRestShape) = 
\frac{1}{\Delta t^2}
\left(
F_{\mathrm{kin}}(\generalizedRestShape) + 
F_{\mathrm{reg}}(\generalizedRestShape)
\right),
\label{eq:rest_shape_optimization}
\end{align}
where $\generalizedRestShapeMin \in \realNumber^{6N-12}$ and $\generalizedRestShapeMax \in \realNumber^{6N-12}$ denote the lower and upper bounds imposed on $\generalizedRestShape$, respectively, and $F_{\mathrm{kin}}(\generalizedRestShape)$ and $F_{\mathrm{reg}}(\generalizedRestShape)$ denote the kinetic energy objective and a regularizer, respectively. Here, we divide $F_{\mathrm{kin}}(\generalizedRestShape)$ and $F_{\mathrm{reg}}(\generalizedRestShape)$ by $\Delta t^2$, anticipating that it will cancel with the $\Delta t^2$ in \eqref{eq:generalized_kinetic_energy1} and \eqref{eq:regularizer}, and noting that this scaling by a constant value has no effect on the rest shape optimization \eqref{eq:rest_shape_optimization}. \mycolor{In addition, note that $\generalizedRestShape$ excludes $\restLength_0$ because the first two root-end vertices are fixed, and thus the edge between them should also be fixed. As such, we do not define the Jacobian with respect to $\restLength_0$ (similar to the fixed variables in Sec. \ref{sec:DER}), and thus $\restLength_0$ is not modified by the rest shape optimization \eqref{eq:rest_shape_optimization}.}

\mycolor{
While one could consider optimizing $\tilde{F}$ in \eqref{eq:rest_shape_optimization} with respect to $\generalizedPos$ given its fewer DOFs than those for $\generalizedRestShape$, and then initializing $\generalizedRestShape$ (which is eventually used to compute DER objectives and forces) with the optimized $\generalizedPos$, this approach makes it difficult to control the system energy (e.g., to avoid stability problems) via box constraints. $\generalizedPos$ is nonlinearly related to multiple components of $\generalizedRestShape$, and thus it is not straightforward to specify box constraints on $\generalizedPos$ that enforce, e.g., $|\restCurvature_{i, 0}| < 1$ to regulate bending energy \eqref{eq:bending_i}. Additionally, computing the Jacobian of the generalized forces with respect to $\generalizedPos$ (i.e., Hessian) is much more complicated compared to those with respect to $\generalizedRestShape$ (see Sec. \ref{sec:rest_shape_optimization_stretching}, Sec. \ref{sec:rest_shape_optimization_bend}, and Sec. \ref{sec:rest_shape_optimization_twist}).
}


\vspace{-4mm}
\subsection{Generalized Kinetic Energy Objective}
\vspace{-2mm}
Considering a single forward timestep, solving \eqref{eq:forward_sim_optimization} determines the generalized force $\generalizedForce_{\mathrm{DER}}$ with  optimal (end of step) generalized positions $\generalizedPos$, i.e., $\generalizedForce_{\mathrm{DER}} = - \nabla_{\generalizedPos} E_{\mathrm{DER}}(\generalizedPos)$. The corresponding optimal (end of step) generalized velocity $\generalizedVel$ is given by 
$
\generalizedVel = 
\generalizedVel^t + 
\dt \generalizedMass^{-1} \generalizedForce_{\mathrm{DER}}$. Thus, the corresponding generalized kinetic energy (summation of the translational kinetic energy for vertices and rotational kinetic energy for edges) can be computed by
\begin{align}
F_{\mathrm{kin}}(\generalizedRestShape) = 
\frac{1}{2}
\norm{\generalizedVel}^2_{\generalizedMass}.
\label{eq:generalized_kinetic_energy0}
\end{align}
Here, we can assume $\generalizedVel^t = 0$ for the static equilibrium case, and therefore the kinetic energy objective \eqref{eq:generalized_kinetic_energy0} can be rewritten as
\begin{align}
F_{\mathrm{kin}}(\generalizedRestShape) = 
\frac{\dt^2}{2}
\norm{\generalizedForce_{\mathrm{DER}}}^2_{\generalizedMass^{-1}}.
\label{eq:generalized_kinetic_energy1}
\end{align}

Considering the Jacobian $\Jacobian_{\mathrm{DER}} (=\frac{\partial \generalizedForce_{\mathrm{DER}}}{\partial \generalizedRestShape}) \in \realNumber^{(4N-8) \times (6N-12)}$ \mycolor{(while treating $\generalizedMass$ as constant because it is aimed to evaluate the norm in the kinetic energy metric at the initial setting)}, the gradient of \eqref{eq:generalized_kinetic_energy1} can be computed as
\begin{align}
\nabla F_{\mathrm{kin}}(\generalizedRestShape) = \dt^2
\Jacobian_{\mathrm{DER}}^T \generalizedMass^{-1} 
\generalizedForce_{\mathrm{DER}},
\label{eq:generalized_kinetic_energy_grad}
\end{align}
and the Hessian can be approximated in the Gauss-Newton style \mycolor{(ignoring the third-order tensors)} by
\begin{align}
\nabla^2 F_{\mathrm{kin}}(\generalizedRestShape) \approx \dt^2
\Jacobian_{\mathrm{DER}}^T \generalizedMass^{-1} \Jacobian_{\mathrm{DER}}.
\label{eq:generalized_kinetic_energy_hess}
\end{align}
Below, we explicitly give the Jacobian for each component.


\vspace{-3mm}
\subsubsection{Inertia}
\vspace{-2mm}
In the static equilibrium case, we can assume $\generalizedPos = \generalizedPos^t$ and $\generalizedVel^t = 0$. As such, fully canceling $\generalizedMass$ (which depends on $\restLength$) in \eqref{eq:inertia_gradient} given the predicted position $\generalizedPos^*$ in \eqref{eq:inertia_objective}, the inertia force $\generalizedForce_{\mathrm{inertia}} = -
\nabla_{\generalizedPos}E_{\mathrm{inertia}}(\generalizedPos)$ can be defined as
\begin{align}
\generalizedForce_{\mathrm{inertia}} =
\generalizedForceExt.
\label{eq:inertia_gradient_static}
\end{align}
This result indicates that the inertia force is independent of $\restLength$, and thus the Jacobian 
$
\frac{\partial \generalizedForce_{\mathrm{inertia}}}
{\partial \generalizedRestShape} = 0$. In addition, \eqref{eq:inertia_gradient_static} is also independent of $\Delta t$, and thus the actual value of $\Delta t$ has no effect on the rest shape optimization \eqref{eq:rest_shape_optimization}. \mycolor{In the rest shape optimization for static equilibrium, a larger norm of the generalized forces, $\norm{\generalizedForce_{\mathrm{inertia}}}_2$, typically requires more significant rest shape changes.}


\vspace{-4mm}
\subsubsection{Stretching}
\label{sec:rest_shape_optimization_stretching}
\vspace{-2mm}
The stretching force of edge $i$ on vertex $i + 1$ is defined as $\generalizedForce_{\mathrm{stretch, i, i + 1}} = -\nabla_{\vertexPos_{i+1}} E_{\mathrm{stretch, i}}(\vertexPos_{i}, \vertexPos_{i+1})$ \eqref{eq:stretch_grad}, and $\generalizedForce_{\mathrm{stretch, i, i}}$ can similarly be defined. Given the dependence of $\generalizedForce_{\mathrm{stretch, i, i + 1}}$ and $\generalizedForce_{\mathrm{stretch, i, i}}$ on $\restLength_i$, we can compute the Jacobian by
\begin{align}
\frac{\partial \generalizedForce_{\mathrm{stretch, i, i+1}}
}{\partial \restLength_i}
= -
\frac{\partial \generalizedForce_{\mathrm{stretch, i, i}}
}{\partial \restLength_i}
= 
\stretchCoef \pi \radius^2 \length_i \restLength_i^{-2} \tangent_i.
\end{align}


\vspace{-4mm}
\subsubsection{Bending}
\label{sec:rest_shape_optimization_bend}
\vspace{-2mm}
We define the bending force according to \eqref{eq:bending_i_grad} by $\generalizedForce_{\mathrm{bend}, i} = -\nabla E_{\mathrm{bend}, i}
(\vertexPos_{i-1}, \edgeAngle_{i-1}, \vertexPos_{i}, \edgeAngle_{i}, \vertexPos_{i+1})$. Given its dependence on $\restLength_{i-1}, \restLength_{i}$, and $\restCurvature_{i}$, the Jacobian can be computed by
\begin{align}
\frac{\partial \generalizedForce_{\mathrm{bend}, i}}{\partial \restLength_{i-1}} =
\frac{\partial \generalizedForce_{\mathrm{bend}, i}}{\partial \restLength_{i}} = 
\left(\frac{\bendCoef \pi \radius^4}{4 (\restLength_{i-1} + \restLength_{i})^2}\right)
\mathbf{J}_{\mathrm{curv}, i}^T (\curvature_i - \restCurvature_i),
\end{align}
and
\begin{align}
\frac{\partial \generalizedForce_{\mathrm{bend}, i}}{\partial \restCurvature_{i}} =
\left(\frac{\bendCoef \pi \radius^4}{4 (\restLength_{i-1} + \restLength_{i})}\right)
\mathbf{J}_{\mathrm{curv}, i}^T.
\end{align}


\vspace{-4mm}
\subsubsection{Twisting}
\label{sec:rest_shape_optimization_twist}
\vspace{-2mm}
Similar to the bending case, we define the twisting force according to \eqref{eq:twist_i_grad} as $\generalizedForce_{\mathrm{twist}, i} = -\nabla E_{\mathrm{twist}, i}
(\vertexPos_{i-1}, \edgeAngle_{i-1}, \vertexPos_{i}, \edgeAngle_{i}, \vertexPos_{i+1})$. Then, given its dependence on $\restLength_{i-1}, \restLength_{i}$, and $\restTwist_{i}$, the Jacobian is computed as
\begin{align}
\frac{\partial \generalizedForce_{\mathrm{twist}, i}}{\partial \restLength_{i-1}} = \frac{\partial \generalizedForce_{\mathrm{twist}, i}}{\partial \restLength_{i}} =
\frac{\twistCoef \pi \radius^4}{(\restLength_{i-1} + \restLength_{i})^2}
(\twist_i - \restTwist_i) \nabla \twist_i,
\end{align}
and
\begin{align}
\frac{\partial \generalizedForce_{\mathrm{twist}, i}}{\partial \restTwist_{i}} =
\frac{\twistCoef \pi \radius^4}{(\restLength_{i-1} + \restLength_{i})} \nabla \twist_i.
\end{align}


\vspace{-4mm}
\subsection{Regularizer for Rest Shape Parameters}
\vspace{-2mm}
Given that the number of DOFs for $\generalizedRestShape$ is larger than for $\generalizedPos$, the system is typically underdetermined, i.e., there are multiple sets of $\generalizedRestShape$ that achieve $F_{\mathrm{kin}}(\generalizedRestShape) = 0$. Among such parameter sets, it is preferable to choose $\generalizedRestShape$ that minimizes deviation from the initial rest shape parameters $\generalizedRestShape_{\mathrm{initial}}$ to avoid introducing stability problems. To this end, we penalize changes in the space of $\generalizedRestShape$ and define the regularizer objective as
\begin{align}
F_{\mathrm{reg}}(\generalizedRestShape) = 
\frac{\dt^2}{2}\regCoef
\norm{\generalizedRestShape - \generalizedRestShape_{\mathrm{initial}}}_2^2,
\label{eq:regularizer}
\end{align}
where $\regCoef$ denotes a \mycolor{positive} regularizer coefficient (we typically set $\regCoef = 10^{-5}$). Although the given system may already be overdetermined (having a unique solution) depending on the initial strand geometry, even in such cases, using the regularizer generally has a positive effect in mitigating overfitting. The previous work \cite{Twigg:2011:OSS:2019406.2019437} proposed using a regularizer based on the force norm, but we found it to be unnecessary because the regularizer in our framework is used to ensure that the system is overdetermined, while the box constraints can restrict the changes in the rest shape parameters in a more controlled way (see Sec. \ref{sec:box_constraints}).

The gradient of \eqref{eq:regularizer} is defined as
\begin{align}
\nabla F_{\mathrm{reg}}(\generalizedRestShape) =
\dt^2 \regCoef
\left(
\generalizedRestShape - \generalizedRestShape_{\mathrm{initial}}
\right),
\label{eq:regularizer_grad}
\end{align}
and the Hessian is given as
\begin{align}
\nabla^2 F_{\mathrm{reg}}(\generalizedRestShape) =
\dt^2 \regCoef \Imat,
\label{eq:regularizer_hess}
\end{align}
where $\Imat$ denotes the Identity matrix.


\vspace{-4mm}
\subsection{Box Constraints for Rest Shape Parameters}
\label{sec:box_constraints}
\vspace{-2mm}
To ensure that the rest shape parameters are within the physically meaningful range and to avoid introducing stability issues into the system, we impose box constraints on these parameters: $\restLength_{\mathrm{min}} \leq \restLength \leq \restLength_{\mathrm{max}}$, $\restCurvature_{\mathrm{min}} \leq \restCurvature \leq \restCurvature_{\mathrm{max}}$, and $\restTwist_{\mathrm{min}} \leq \restTwist \leq \restTwist_{\mathrm{max}}$.

Given the initial rest length $\restLength_{\mathrm{initial}} \ (> 0)$, we define the lower and upper bounds as
\begin{align}
\restLength_{\mathrm{min}} = a_{\mathrm{min}} \restLength_{\mathrm{initial}},
\quad
\restLength_{\mathrm{max}} = a_{\mathrm{max}} \restLength_{\mathrm{initial}},
\label{eq:rest_len_min_max}
\end{align}
where $a_{\mathrm{min}}$ and $a_{\mathrm{max}}$ are scalars, and we use $a_{\mathrm{min}} = 0.1$ and $a_{\mathrm{max}} = 1.1$ (which are determined through our experiments).

Considering the initial rest curvature $\restCurvature_{\mathrm{initial}}$ and rest twist $\restTwist_{\mathrm{initial}}$ (where both of them can be 0 in contrast to $\restLength_{\mathrm{initial}}$), we define their lower and upper bounds as
\begin{align}
\restCurvature_{\mathrm{min}} &= \restCurvature_{\mathrm{initial}} - \restCurvatureDelta,
\quad
\restCurvature_{\mathrm{max}} = \restCurvature_{\mathrm{initial}} + \restCurvatureDelta,
\label{eq:rest_curvature_min_max}
\\
\restTwist_{\mathrm{min}} &= \restTwist_{\mathrm{initial}} - \restTwistDelta,
\quad
\restTwist_{\mathrm{max}} = \restTwist_{\mathrm{initial}} + \restTwistDelta,
\label{eq:rest_twist_min_max}
\end{align}
where column vectors $\restCurvatureDelta \in \realNumber^{4N-8}$ and $\restTwistDelta \in \realNumber^{N-2}$ denote the permitted deviation from $\restCurvature_{\mathrm{initial}}$ and $\restTwist_{\mathrm{initial}}$, respectively.

Given the turning angle $\turningAngle_i$ at vertex $i$ \cite{Bergou2008}, we can derive the relation between the norm of the curvature and turning angle as $\frac{1}{2}\norm{\curvature_i}_2^2 = \left(2\tan \frac{\turningAngle_i}{2}\right)^2$. Thus, if we have the change of the turning angle $\Delta \turningAngle$ at $\turningAngle_i = 0$ with the corresponding change of the curvature $\Delta \kappa$ (assuming $\Delta \kappa = \Delta \curvature_{i, 0} = \Delta \curvature_{i, 1} = \Delta \curvature_{i, 2} = \Delta \curvature_{i, 3}$ for simplicity), we get $\frac{1}{2}(4\Delta \kappa)^2 = \left(2\tan \frac{\Delta \turningAngle}{2}\right)^2$, and thus $|\Delta \kappa| = \sqrt{2}|\tan \frac{\Delta \turningAngle}{2}|$. This relation indicates that, e.g., allowing a maximal change of the turning angle of $\frac{\pi}{2}$ (i.e., $\Delta \turningAngle = \frac{\pi}{2}$) leads to a change of the curvature  of $\Delta \kappa = \sqrt{2}$. Based on this analysis for bending, we define $\restCurvatureDelta$ as
\begin{align}
\restCurvatureDelta 
= \Delta \kappa \evec_{\restCurvatureDelta} 
= \sqrt{2} \evec_{\restCurvatureDelta},
\label{eq:b}
\end{align}
where $\evec_{\restCurvatureDelta}$ denotes a vector of all ones with the same length as $\restCurvatureDelta$. For twisting, we permit a maximal change of twist of $\Delta m = \frac{\pi}{8}$, and thus we define $\restTwistDelta$ as
\begin{align}
\restTwistDelta
= \Delta m \evec_{\restTwistDelta}
= \frac{\pi}{8} \evec_{\restTwistDelta},
\label{eq:d}
\end{align}
where $\evec_{\restTwistDelta}$ denotes a vector of all ones with the same length as $\restTwistDelta$. Since we chose $\Delta \curvature$ in \eqref{eq:b} and $\Delta m$ in \eqref{eq:d} empirically based on our analysis and numerical experiments with reasonable material parameters (e.g., $\stretchCoef = 10^{8} \ \mathrm{kg/(m\cdot s^{2})}$, $\bendCoef = 10^{8} \ \mathrm{kg/(m\cdot s^{2})}$, and $\twistCoef = 10^{8} \ \mathrm{kg/(m\cdot s^{2})}$), it would be necessary to adjust these values for much softer materials (Sec. \ref{sec:box_constraint_evaluation}).


\vspace{-4mm}
\subsection{Gauss-Newton Solver with Penalty Method}
\vspace{-2mm}
Given the least-squares-style nonlinear objectives arising in our problem, we aim to solve the optimization via the GN method \cite{NoceWrig06}. While enforcing box constraints within Newton-type optimizers is possible \cite{Takahashi2021}, it requires accounting for the box constraints when solving inner problems, which leads to an additional computational cost compared to unconstrained linear solves. Considering that our problems are relatively stiff (albeit yielding sparse systems), instead of handling box constraints in the inner problems, we prefer to convert the box-constrained minimization into an unconstrained one via the penalty method \cite{NoceWrig06}, thus ensuring unconstrained inner linear systems for greater efficiency.

Specifically, we reformulate the box-constrained minimization \eqref{eq:rest_shape_optimization} into the following unconstrained minimization by converting the box constraints into a penalty objective $F_{\mathrm{box}}(\generalizedRestShape)$ \cite{NoceWrig06}:
\begin{align}
\generalizedRestShape = 
\argmin_{
\generalizedRestShape
} F(\generalizedRestShape),
\
F(\generalizedRestShape) = 
\frac{1}{\Delta t^2}
\left(
F_{\mathrm{kin}}(\generalizedRestShape) + 
F_{\mathrm{reg}}(\generalizedRestShape) +
F_{\mathrm{box}}(\generalizedRestShape)
\right),
\label{eq:rest_shape_optimization_unconstrained}
\end{align}
where we define $F_{\mathrm{box}}(\generalizedRestShape)$ as
\begin{align}
F_{\mathrm{box}}(\generalizedRestShape) = 
\frac{\dt^2}{2}\penaltyCoef
\left(
\norm{
\max(\generalizedRestShape - \generalizedRestShapeMax, 0)
}_2^2 +
\norm{
\max(\generalizedRestShapeMin - \generalizedRestShape, 0)
}_2^2
\right),
\label{eq:box_penalty}
\end{align}
with $\penaltyCoef$ denoting a penalty parameter (we set $\penaltyCoef = 10^{6}$). \mycolor{Note that, similar to the original rest shape optimization \eqref{eq:rest_shape_optimization},  $\dt$ in \eqref{eq:rest_shape_optimization_unconstrained} cancels completely with \eqref{eq:generalized_kinetic_energy1}, \eqref{eq:regularizer}, \eqref{eq:box_penalty}, and therefore $\dt$ has no effect on achieving static equilibrium via rest shape optimization.}

The gradient of \eqref{eq:box_penalty} is given by
\begin{align}
\nabla F_{\mathrm{box}}(\generalizedRestShape) = \dt^2 \penaltyCoef
\left(
\max(\generalizedRestShape - \generalizedRestShapeMax, 0) +
\max(\generalizedRestShapeMin - \generalizedRestShape, 0)
\right),
\label{eq:box_grad}
\end{align}
and the Hessian is computed by
\begin{align}
\nabla^2 F_{\mathrm{box}}(\generalizedRestShape) = \dt^2 \penaltyCoef
\left(
\Hmat_{\mathrm{max}} + \Hmat_{\mathrm{min}}
\right),
\label{eq:box_hess}
\end{align}
where $\Hmat_{\mathrm{max}} \in \realNumber^{(6N-12) \times (6N-12)}$ is a diagonal matrix, with diagonal elements given by $\mathrm{diag}(\Hmat_{\mathrm{max}}) = [H(\generalizedRestShape_0 - \generalizedRestShape_{\mathrm{max}, 0}), \ldots, H(\generalizedRestShape_{6N-11} - \generalizedRestShape_{\mathrm{max}, 6N-11})]^T$ using the Heaviside step function $H(\cdot)$ with $H(0) = 0$. We define $\Hmat_{\mathrm{min}}$ analogously.

Algorithm \ref{alg:gauss_newton} shows our rest shape optimization algorithm with the GN solver. To accelerate its convergence via warm starting and to reduce the possibility of falling into suboptimal local minima, we initialize $\generalizedRestShape$ from the initial $\length$, $\curvature$, and $\twist$. The inner linear system is guaranteed to be symmetric positive definite (SPD) due to the GN-style Hessian approximation and the regularizer. Given that the system is relatively stiff yet sparse, we solve it using a sparse Cholesky-based direct solver. To ensure a decrease of the objective, we employ a back-tracking line search \cite{NoceWrig06}. In practice, as we enforce the box constraints via the penalty, the rest shape parameters $\generalizedRestShape$ can slightly violate their bounds. Given these relatively small violations, if necessary, we can project $\generalizedRestShape$ back into the valid range in each iteration (although this treatment was not needed in any of our examples). We terminate the GN iterations when $\norm{\Delta \generalizedRestShape}_2$ becomes smaller than a threshold $\epsilon \ (= 10^{-5})$. In addition, to ensure that the GN solver ultimately terminates, we halt the iteration if it exceeds $500$ iterations or the back-tracking line search fails.

\vspace{-2mm}
\begin{algorithm}
\caption{Rest Shape Optimization with Gauss-Newton}
\label{alg:gauss_newton}
\begin{algorithmic}[1]
\STATE $k=0$
\STATE Initialize the generalized positions $\generalizedPos$, density $\rho$, radius $\radius$, and material stiffness coefficients $\stretchCoef, \bendCoef$, and $\twistCoef$
\STATE Initialize the unit tangent vector $\tangent$ and length $\length$, and set $\restLength = \length$
\STATE Initialize the generalized mass matrix $\generalizedMass$
\STATE Initialize the reference frames and update material frames
\STATE Compute the curvature $\curvature$ and twist $\twist$, and $\restCurvature = \curvature$ and $\restTwist = \twist$
\STATE Set $\restLength_{\mathrm{initial}} = \restLength$, $\restCurvature_{\mathrm{initial}} = \restCurvature$, and $\restTwist_{\mathrm{initial}} = \restTwist$
\STATE Initialize 
$\restLength_{\mathrm{min}}, \restLength_{\mathrm{max}}$ 
with \eqref{eq:rest_len_min_max}, 
$\restCurvature_{\mathrm{min}}, \restCurvature_{\mathrm{max}}$
with \eqref{eq:rest_curvature_min_max},
and
$\restTwist_{\mathrm{min}}, \restTwist_{\mathrm{max}}$
with \eqref{eq:rest_twist_min_max}
\STATE \textbf{do}
\STATE \quad Compute gradient $\nabla F(\generalizedRestShape^{k})$ with \eqref{eq:generalized_kinetic_energy_grad}, \eqref{eq:regularizer_grad}, and \eqref{eq:box_grad}
\STATE \quad Approximate Hessian $\nabla^2 F(\generalizedRestShape^{k})$ with 
\eqref{eq:generalized_kinetic_energy_hess}, \eqref{eq:regularizer_hess}, and \eqref{eq:box_hess}
\STATE \quad Compute $\Delta \generalizedRestShape^{k+1}$ by solving $(\nabla^2 F(\generalizedRestShape^{k})) \Delta \generalizedRestShape^{k+1} = -\nabla F(\generalizedRestShape^{k})$ with a sparse Cholesky-based direct solver
\STATE \quad Compute the step length $\gamma$ using back-tracking line search with \eqref{eq:generalized_kinetic_energy1}, \eqref{eq:regularizer}, and \eqref{eq:box_penalty} 
\STATE \quad Update the rest shape parameters by $\generalizedRestShape^{k+1} = \generalizedRestShape^{k} + \gamma \Delta \generalizedRestShape^{k+1}$
\STATE \quad Optionally, clamp $\generalizedRestShape^{k+1}$ into the valid range
\STATE \quad $k = k + 1$
\STATE \textbf{while} $\norm{\Delta \generalizedRestShape^{k+1}}_2 > \epsilon$
\end{algorithmic}
\end{algorithm}
\vspace{-2mm}


\vspace{-4mm}
\subsection{Simplification with Zero Rest Twist}
\vspace{-2mm}
While capturing the twisting effects of elastic strands is essential for visual fidelity, it is relatively rare to exactly specify initial twist or edge angles when designing a strand geometry, e.g., for hair assets, because edge angles have almost no effect on the visualization result at the static state, and capturing and specifying edge angle information is challenging \cite{Luo2013hair,Hu2015hair}. In addition, some elastic rod simulators do not necessarily support edge angles \cite{Selle2008mass,Umetani2014rod,Shi2023curls} and it is generally difficult for humans to intuit how the edge angles influence the strand dynamics for simulators that do consider the edge angles \cite{Bergou2008,Bergou:2010:DVT:1778765.1778853,Kugelstadt2016rod,Soler2018rod}. As a result, it is common to specify only vertex positions and initialize edge angles $\edgeAngle$ to zero. Given such zero edge angles, which also yield zero initial twist $\twist$, setting the rest twist $\restTwist$ to zero completely eliminates the twist energy \eqref{eq:twist_i} and its gradient \eqref{eq:twist_i_grad}. As such, we can potentially exclude $\restTwist$ from $\generalizedRestShape$, leading to an effective DOF count of $(5N-10)$ for $\generalizedRestShape$ and thus reducing the computational cost for rest shape optimization. 

In practice, rest shape optimization without the rest twist still works well (i.e., achieves a stable static equilibrium) if the bending of the strands introduces only a limited amount of twisting given the tight coupling of bending and twisting in the DER formulation. Such cases do occur, so we will evaluate and compare the rest shape optimization with and without rest twist in terms of performance and quality \mycolor{(see Sec. \ref{sec:stress} and Sec. \ref{sec:hair_like})}.


\vspace{-4mm}
\subsection{\mycolor{Discussion of Norm Formulations}}
\label{sec:kinetic_energy_vs_force_norm}
\vspace{-2mm}
Eliding unnecessary coefficients for a simpler analysis, the objective based on the $L^2$ norm of \mycolor{the generalized force $\generalizedForce$ is given by $\norm{\generalizedForce}_2^2$ \cite{Twigg:2011:OSS:2019406.2019437}} whereas our kinetic-energy-based objective is $\norm{\generalizedForce}_{\generalizedMass^{-1}}^2$ due to \eqref{eq:generalized_kinetic_energy1}. As an example, consider the inertia force \eqref{eq:inertia_gradient_static} due to \mycolor{generalized gravity $\gravity$ since the inertia is the unique trigger for non-zero elastic forces (which are expected to cancel the inertia)} in our framework. In this case, $\generalizedForce = \generalizedMass \gravity$, and thus we get $\gravity^T \generalizedMass^2 \gravity$ for the force-norm-based objective versus $\gravity^T \generalizedMass \gravity$ for ours. As such, when mass/inertia of the strands is too small or large, squaring $\generalizedMass$ can negatively impact the numerical conditioning of the system. \mycolor{In practice, Gauss-Newton optimization on $\norm{\generalizedForce}_2^2$ through $(\Jacobian^T \Jacobian) \Delta \generalizedRestShape = -\Jacobian^T \generalizedForce$ (where $\Jacobian$ denotes the Jacobian of $\generalizedForce$ with respect to $\generalizedRestShape$) fails to reach sufficiently small $\norm{\generalizedForce}_2^2$ because $\Jacobian^T \Jacobian$ has entries significantly different in their magnitude and thus is numerically ill-conditioned, neglecting some force components. Consequently,} extended precision would be required for the optimization to consistently succeed \cite{Twigg:2011:OSS:2019406.2019437}. \mycolor{Additionally, it is worth noting that our formulation properly accounts for the energy contributions from each component (unlike simple scaling and dimension analysis approaches that merely mitigate numerical precision problems) and is essential for DER as it involves variables in different dimensions (vertex positions and edge angles).}


\vspace{-4mm}
\section{Results and Discussions}
\vspace{-2mm}
We implemented our method in C++20 with double-precision floating-point for scalars. When multiple strands are involved, we parallelize forward simulation and inverse problems with OpenMP, processing each strand in parallel. We executed the examples on a desktop machine with an Intel Core i7-9700 (8 cores) with 16GB RAM. We use 60 frames per second with a single simulation step except for Figure \ref{fig:teaser} and Figure \ref{fig:hairs}, where we used five steps. We include the rest twist in the optimization unless otherwise mentioned. We set $\rho = 10^3 \ \mathrm{kg/m^3}$, $\radius = 10^{-3} \ \mathrm{m}$, $\stretchCoef = 10^{8} \ \mathrm{kg/(m\cdot s^{2})}$, $\bendCoef = 10^{8} \ \mathrm{kg/(m\cdot s^{2})}$, and $\twistCoef = 10^{8} \ \mathrm{kg/(m\cdot s^{2})}$ unless otherwise specified. For visualization purposes, we enlarge the vertices and edges. The first two black vertices and cyan edge indicate the ``fixed'' vertices and edge, respectively, while the red vertices and blue and white edges represent movable vertices and edges with active DOFs, respectively (except for \mycolor{Figures \ref{fig:teaser}, \ref{fig:coil}, and \ref{fig:hairs}}). In addition, we render the material frames at the centers of the corresponding edges as green and magenta lines whose lengths are the same as the rest lengths of the edges.  


\begin{figure}[tb]
\captionsetup[subfigure]{aboveskip=0mm,belowskip=0mm,font=small}
\centering
\begin{subfigure}[b]{0.333\linewidth}
    \adjincludegraphics[trim={{0.45\width} {0.0\height} {0.45\width} {0.0\height}}, clip, width=0.333\linewidth]
    {figures/\figDir/vertical_00_5e3_wo_rso_sim.png}%
    \adjincludegraphics[trim={{0.45\width} {0.0\height} {0.45\width} {0.0\height}}, clip, width=0.333\linewidth]
    {figures/\figDir/vertical_01_5e3_w_rso_sim.png}%
    \adjincludegraphics[trim={{0.45\width} {0.0\height} {0.45\width} {0.0\height}}, clip, width=0.333\linewidth]
    {figures/\figDir/vertical_02_5e3_w_rso_sim_no_grav.png}
\caption{$\stretchCoef = 5\times 10^3$}
\end{subfigure}%
\begin{subfigure}[b]{0.333\linewidth}
    \adjincludegraphics[trim={{0.45\width} {0.0\height} {0.45\width} {0.0\height}}, clip, width=0.333\linewidth]
    {figures/\figDir/vertical_03_5e4_wo_rso_sim.png}%
    \adjincludegraphics[trim={{0.45\width} {0.0\height} {0.45\width} {0.0\height}}, clip, width=0.333\linewidth]
    {figures/\figDir/vertical_04_5e4_w_rso_sim.png}%
    \adjincludegraphics[trim={{0.45\width} {0.0\height} {0.45\width} {0.0\height}}, clip, width=0.333\linewidth]
    {figures/\figDir/vertical_05_5e4_w_rso_sim_no_grav.png}
\caption{$\stretchCoef = 5\times 10^4$}
\end{subfigure}%
\begin{subfigure}[b]{0.333\linewidth}
    \adjincludegraphics[trim={{0.45\width} {0.0\height} {0.45\width} {0.0\height}}, clip, width=0.333\linewidth]
    {figures/\figDir/vertical_06_5e5_wo_rso_sim.png}%
    \adjincludegraphics[trim={{0.45\width} {0.0\height} {0.45\width} {0.0\height}}, clip, width=0.333\linewidth]
    {figures/\figDir/vertical_07_5e5_w_rso_sim.png}%
    \adjincludegraphics[trim={{0.45\width} {0.0\height} {0.45\width} {0.0\height}}, clip, width=0.333\linewidth]
    {figures/\figDir/vertical_08_5e5_w_rso_sim_no_grav.png}
\caption{$\stretchCoef = 5\times 10^5$}
\end{subfigure}%
\vspace{-4mm}
\caption{
Evaluation with a single vertical strand. For each $\stretchCoef$ trio: (left) simulation results with naive initialization, (middle) with rest shape optimization, and (right) with rest shape optimization and zero gravity during simulation. Our rest shape optimizer modifies the rest length more significantly for softer materials to completely cancel gravity and thus achieve static equilibrium.
\vspace{-2mm}
}
\label{fig:vertical}
\end{figure}

\vspace{-4mm}
\subsection{Single Vertical Strand}
\vspace{-2mm}
To illustrate the effectiveness of our rest shape optimization, we first experiment with a single vertical elastic strand \mycolor{(whose length is $1 \mathrm{m}$)}, uniformly discretized with 20 vertices, as shown in Figure \ref{fig:vertical}. With three different stretch coefficients $\stretchCoef$ ($5\times 10^3, 5\times 10^4$, and $5\times 10^5$), we compare simulation results with the following three cases:
\begin{enumerate}
\item Naive initialization: initialization without our rest shape optimization;
\item With rest shape optimization;
\item With rest shape optimization, and zero gravity during simulation.
\end{enumerate}

Due to the vertical arrangement of the strands, our method modifies the rest length only to counteract gravity without changing the rest curvature or rest twist, even though both are included in the rest shape optimization. Without the rest shape optimization the strand sags due to gravity, whereas our method enables the strand to achieve static equilibrium under all three different stiffness settings, modifying the rest length more significantly for softer materials to completely cancel the gravity force. Disabling gravity after the rest shape optimization reveals the modified rest lengths, which are shorter than the original ones.


\begin{figure}[tb]
\captionsetup[subfigure]{aboveskip=0mm,belowskip=0mm,font=small}
\centering
\adjincludegraphics[trim={{0.0\width} {0.0\height} {0.0\width} {0.0\height}}, clip, width=1.0\linewidth]
{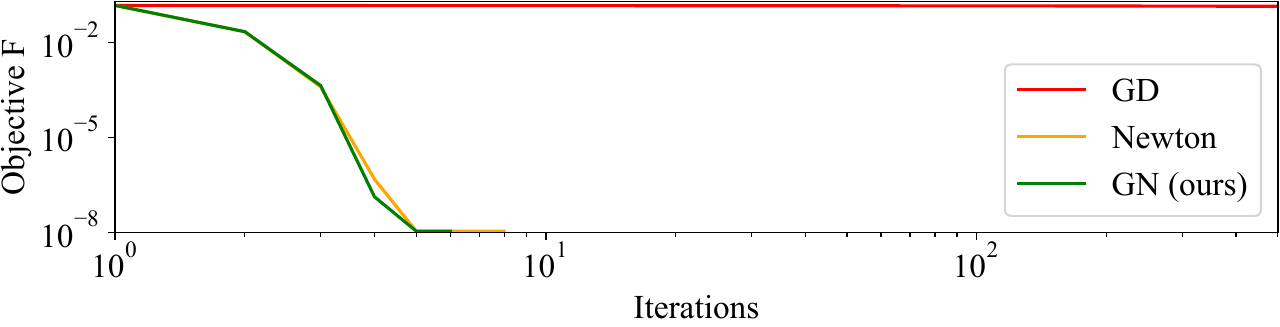}%
\vspace{-4mm}
\caption{
Convergence profile for GD, Newton, and GN (ours). Both Newton and GN quickly converge while GD requires many more iterations.
\vspace{-6mm}
}
\label{fig:optimizer}
\end{figure}

\vspace{-4mm}
\subsection{Nonlinear Optimizer Comparisons}
\vspace{-2mm}
To justify our choice for the nonlinear optimizer, we again experiment with the vertical strand scenario (similar to Fig. \ref{fig:vertical}), using 500 vertices and $\stretchCoef = 10^4$. We compare the following schemes:
\begin{enumerate}
\item GD: gradient descent;
\item Newton: Newton's method;
\item GN (ours): Gauss-Newton method.
\end{enumerate}
For Newton's method, we compute the exact Hessian via the third-order tensors for \eqref{eq:rest_shape_optimization_unconstrained} with the stretching force (since bending and twisting will not be introduced in this setting) and ensure an SPD system via Hessian projection per edge. Figure \ref{fig:optimizer} shows a convergence profile of the objective over optimization iterations. 

The back-tracking line search leads GD to require tiny steps due to the stiff system, and GD fails to sufficiently decrease the objective within 500 iterations. \mycolor{Due to the one-dimensional strand structure, a sparse Cholesky solver with matrix reordering can solve the system (linear solve is required only for Newton-type optimizers, but not for GD) with an almost linear cost and is sufficiently fast (per-iteration cost for GN is less than $2\times$ of GD), and thus GN is much faster (GD took $10.7$ s with 500 iterations whereas GN took $0.11$ s with 6 iterations). Given GN's low per-iteration cost, it is likely to outperform quasi-Newton methods.} With Newton's method, we do not observe a particular benefit on convergence over GN while computing the exact Hessian introduces an additional cost and significant implementation complexity associated with the third-order tensors. These observations make GN most suitable as our optimizer.


\begin{figure}[tb]
\captionsetup[subfigure]{aboveskip=0mm,belowskip=0mm,font=small}
\centering
\begin{subfigure}[b]{1.0\linewidth}
    \adjincludegraphics[trim={{0.15\width} {0.0\height} {0.15\width} {0.1\height}}, clip, width=0.333\linewidth]
    {figures/\figDir/horizontal_09_10e7_wo_rso_sim.png}%
    \adjincludegraphics[trim={{0.15\width} {0.0\height} {0.15\width} {0.1\height}}, clip, width=0.333\linewidth]
    {figures/\figDir/horizontal_10_10e7_w_rso_sim.png}%
    \adjincludegraphics[trim={{0.15\width} {0.0\height} {0.15\width} {0.1\height}}, clip, width=0.333\linewidth]
    {figures/\figDir/horizontal_11_10e7_w_rso_sim_no_grav.png}
    \caption{$\bendCoef = 10^{7}$}
\end{subfigure}
\begin{subfigure}[b]{1.0\linewidth}
    \adjincludegraphics[trim={{0.15\width} {0.0\height} {0.15\width} {0.1\height}}, clip, width=0.333\linewidth]
    {figures/\figDir/horizontal_00_10e8_wo_rso_sim.png}%
    \adjincludegraphics[trim={{0.15\width} {0.0\height} {0.15\width} {0.1\height}}, clip, width=0.333\linewidth]
    {figures/\figDir/horizontal_01_10e8_w_rso_sim.png}%
    \adjincludegraphics[trim={{0.15\width} {0.0\height} {0.15\width} {0.1\height}}, clip, width=0.333\linewidth]
    {figures/\figDir/horizontal_02_10e8_w_rso_sim_no_grav.png}
    \caption{$\bendCoef = 10^{8}$}
\end{subfigure}
\begin{subfigure}[b]{1.0\linewidth}
    \adjincludegraphics[trim={{0.15\width} {0.0\height} {0.15\width} {0.1\height}}, clip, width=0.333\linewidth]
    {figures/\figDir/horizontal_03_10e9_wo_rso_sim.png}%
    \adjincludegraphics[trim={{0.15\width} {0.0\height} {0.15\width} {0.1\height}}, clip, width=0.333\linewidth]
    {figures/\figDir/horizontal_04_10e9_w_rso_sim.png}%
    \adjincludegraphics[trim={{0.15\width} {0.0\height} {0.15\width} {0.1\height}}, clip, width=0.333\linewidth]
    {figures/\figDir/horizontal_05_10e9_w_rso_sim_no_grav.png}
    \caption{$\bendCoef = 10^{9}$}
\end{subfigure}
\begin{subfigure}[b]{1.0\linewidth}
    \adjincludegraphics[trim={{0.15\width} {0.0\height} {0.15\width} {0.1\height}}, clip, width=0.333\linewidth]
    {figures/\figDir/horizontal_06_10e10_wo_rso_sim.png}%
    \adjincludegraphics[trim={{0.15\width} {0.0\height} {0.15\width} {0.1\height}}, clip, width=0.333\linewidth]
    {figures/\figDir/horizontal_07_10e10_w_rso_sim.png}%
    \adjincludegraphics[trim={{0.15\width} {0.0\height} {0.15\width} {0.1\height}}, clip, width=0.333\linewidth]
    {figures/\figDir/horizontal_08_10e10_w_rso_sim_no_grav.png}
    \caption{$\bendCoef = 10^{10}$}
\end{subfigure}%
\vspace{-4mm}
\caption{
Evaluation with a single horizontal strand. For each $\bendCoef$ trio: (left) simulation results with naive initialization, (middle) with rest shape optimization, and (right) with rest shape optimization and zero gravity during simulation. Our optimizer modifies the rest curvature more significantly for softer strands to achieve static equilibrium \mycolor{(b), (c), and (d) while it fails for too soft material (a)}.
\vspace{-6mm}
}
\label{fig:horizontal}
\end{figure}

\vspace{-4mm}
\subsection{Single Horizontal Strand}
\vspace{-2mm}
Next, we test with a horizontal strand uniformly discretized with 20 vertices, as shown in Figure \ref{fig:horizontal}, \mycolor{and used $\Delta \curvature = 10$ to allow larger curvature changes.} With \mycolor{four} different bending coefficients $\bendCoef$ (\mycolor{$10^7, 10^8, 10^9$}, and $10^{10}$), we compare the three cases:
\begin{enumerate}
\item Naive initialization;
\item With rest shape optimization;
\item With rest shape optimization, and zero gravity during simulation.
\end{enumerate}

Due to the horizontal arrangement of the strands, our rest shape optimizer modifies the rest curvatures only (but not rest lengths or rest twists). \mycolor{While we can achieve static equilibrium with the three stiffer strands, with our optimizer modifying the rest curvature more significantly for the softer materials, the softest strand fails because it requires much larger rest curvature changes (which are not permitted by the box constraints).} While increasing the stiffness is one way to reduce the sagging at equilibrium, this approach can change the strand dynamics and our perception of the material.


\begin{figure}[tb]
\captionsetup[subfigure]{aboveskip=0mm,belowskip=0mm,font=small}
\centering
\begin{subfigure}[b]{0.5\linewidth}
    \adjincludegraphics[trim={{0.15\width} {0.2\height} {0.15\width} {0.0\height}}, clip, width=0.5\linewidth]
    {figures/\figDir/load_00_rso.png}%
    \adjincludegraphics[trim={{0.15\width} {0.2\height} {0.15\width} {0.0\height}}, clip, width=0.5\linewidth]
    {figures/\figDir/load_01_rso_no_grav.png}%
    \caption{Without a load}
\end{subfigure}%
\begin{subfigure}[b]{0.5\linewidth}
    \adjincludegraphics[trim={{0.15\width} {0.2\height} {0.15\width} {0.0\height}}, clip, width=0.5\linewidth]
    {figures/\figDir/load_02_rso_load.png}%
    \adjincludegraphics[trim={{0.15\width} {0.2\height} {0.15\width} {0.0\height}}, clip, width=0.5\linewidth]
    {figures/\figDir/load_03_rso_load_no_grav.png}%
    \caption{With a load}
\end{subfigure}%
\vspace{-4mm}
\caption{
Test with a horizontal strand and an additional load (as illustrated with an orange arrow) using our rest shape optimization. For each case, we show the simulation with gravity (left) and without gravity (right). To counteract the additional load, our rest shape optimizer modifies the rest curvature more significantly.
\vspace{-4mm}
}
\label{fig:load}
\end{figure}

\vspace{-4mm}
\subsection{External Load}
\vspace{-2mm}
We also introduce an additional external load (which is a force $10\times$ larger than  gravity) exactly on the tail-end vertex of the horizontal strand, as shown in Figure \ref{fig:load}. While static equilibrium can be achieved even with the load, our method modifies the rest curvature further to cancel the force due to the load.


\begin{figure}[tb]
\captionsetup[subfigure]{aboveskip=0mm,belowskip=0mm,font=small}
\centering
\begin{subfigure}[b]{0.5\linewidth}
    \adjincludegraphics[trim={{0.2\width} {0.3\height} {0.15\width} {0.3\height}}, clip, width=1.0\linewidth]
    {figures/\figDir/04_stress_w_twist.png}
    \caption{With rest twist optimization}
\end{subfigure}%
\begin{subfigure}[b]{0.5\linewidth}
    \adjincludegraphics[trim={{0.2\width} {0.3\height} {0.15\width} {0.3\height}}, clip, width=1.0\linewidth]
    {figures/\figDir/04_stress_wo_twist.png}
    \caption{Without rest twist optimization}
\end{subfigure}%
\vspace{-4mm}
\caption{
Stress test with a horizontal strand discretized with $1,000$ vertices. Both approaches successfully achieve static equilibrium.
\vspace{-6mm}
}
\label{fig:stress}
\end{figure}

\vspace{-4mm}
\subsection{Stress Test}
\label{sec:stress}
\vspace{-2mm}
To demonstrate the capabilities of our method with high-resolution strands, we experiment with a horizontal strand discretized with $1,000$ vertices, as shown in Figure \ref{fig:stress} (notably, hair strands are more commonly discretized with up to $100$ vertices \cite{Shi2023curls}). Given the horizontal structure of the strand, bending does not introduce twisting, and thus we can safely exclude the rest twist from the rest shape optimization. Our method achieved  static equilibrium in both cases, i.e., including and excluding rest twist for the rest shape optimization. The computation times were $0.087$ s and $0.072$ s which are proportional to the number of DOFs, $(6N-12)$ and $(5N-10)$, respectively.


\begin{figure}[tb]
\captionsetup[subfigure]{aboveskip=0mm,belowskip=0mm,font=small}
\centering
\begin{subfigure}[b]{0.333\linewidth}
    \adjincludegraphics[trim={{0.35\width} {0.0\height} {0.4\width} {0.0\height}}, clip, width=0.333\linewidth]
    {figures/\figDir/hair_00_w_twist_0001.png}%
    \adjincludegraphics[trim={{0.35\width} {0.0\height} {0.4\width} {0.0\height}}, clip, width=0.333\linewidth]
    {figures/\figDir/hair_00_w_twist_0050.png}%
    \adjincludegraphics[trim={{0.35\width} {0.0\height} {0.4\width} {0.0\height}}, clip, width=0.333\linewidth]
    {figures/\figDir/hair_00_w_twist_0300.png}%
    \caption{Ours with non-\\zero rest twist}
\end{subfigure}%
\begin{subfigure}[b]{0.333\linewidth}
    \adjincludegraphics[trim={{0.35\width} {0.0\height} {0.4\width} {0.0\height}}, clip, width=0.333\linewidth]
    {figures/\figDir/hair_01_wo_twist_0001.png}%
    \adjincludegraphics[trim={{0.35\width} {0.0\height} {0.4\width} {0.0\height}}, clip, width=0.333\linewidth]
    {figures/\figDir/hair_01_wo_twist_0050.png}%
    \adjincludegraphics[trim={{0.35\width} {0.0\height} {0.4\width} {0.0\height}}, clip, width=0.333\linewidth]
    {figures/\figDir/hair_01_wo_twist_0300.png}%
    \caption{Ours with zero\\ rest twist}
\end{subfigure}%
\begin{subfigure}[b]{0.333\linewidth}
    \adjincludegraphics[trim={{0.35\width} {0.0\height} {0.4\width} {0.0\height}}, clip, width=0.333\linewidth]
    {figures/\figDir/hair_02_two_0001.png}%
    \adjincludegraphics[trim={{0.35\width} {0.0\height} {0.4\width} {0.0\height}}, clip, width=0.333\linewidth]
    {figures/\figDir/hair_02_two_0050.png}%
    \adjincludegraphics[trim={{0.35\width} {0.0\height} {0.4\width} {0.0\height}}, clip, width=0.333\linewidth]
    {figures/\figDir/hair_02_two_0300.png}%
    \caption{Global-local\\ initialization}
\end{subfigure}%
\vspace{-4mm}
\caption{
An experiment with a single hair-like strand. Enforcing zero rest twist (b) leads to a relatively unstable static equilibrium initially, causing flipping that yields a kink. Global-local initialization of DER (c) fails to achieve static equilibrium.
\vspace{-6mm}
}
\label{fig:hair_strand}
\end{figure}

\vspace{-4mm}
\subsection{Single Hair-Like Strand}
\label{sec:hair_like}
\vspace{-2mm}
Next, we experiment with a single hair-like strand (which can introduce non-negligible torsion effects, unlike vertical and horizontal strands), as shown in Figure \ref{fig:hair_strand}, to evaluate our method including/excluding rest twist in the optimization, along with the global-local initialization \cite{Hsu2022sag}. Specifically, we compare the following schemes:
\begin{enumerate}
\item Ours with non-zero rest twist: we include rest twist in the optimization, which generally leads to non-zero rest twist;
\item Ours with zero rest twist: we exclude rest twist from the optimization, setting rest twist to zero;
\item Global-local initialization: \cite{Hsu2022sag}.
\end{enumerate}
Since directly applying the global-local initialization \cite{Hsu2022sag} to DER is not possible, as discussed in Sec. \ref{sec:global_local}, we use the following two simplifications. First, we use the minimal stencil of the bending and twisting as a unified single force element to form a global linear system, assuming that the force element is linear/angular momentum preserving. Second, we first determine rest length from the stretch force without considering the bending and twisting forces to avoid transforming the local step into a global problem assuming relatively small changes in the rest length due to the large $\stretchCoef$. Then, we optimize the rest curvature and rest twist using the rest length and local-element forces (computed by the global linear solve) in a least-squares way.

The strand with non-zero rest twist achieves static equilibrium and retains the same shape. By contrast, while the strand with zero rest twist can achieve static equilibrium at first, it is relatively unstable and flips in its middle part to reach a more stable static equilibrium. This is because the bending forces need to cancel any torques by themselves (as the twist forces are zero by construction at first) leading to significant changes in the rest curvatures and thus the unstable static equilibrium. Due to the necessary simplifications, the global-local initialization \cite{Hsu2022sag} failed to achieve static equilibrium (although the global-local initialization, as described above, was able to achieve static equilibrium for perfectly vertical or horizontal strands in our experiments).


\begin{figure}[tb]
\captionsetup[subfigure]{aboveskip=0mm,belowskip=0mm,font=small}
\centering
\begin{subfigure}[b]{0.333\linewidth}
    \adjincludegraphics[trim={{0.4\width} {0.0\height} {0.4\width} {0.0\height}}, clip, width=0.333\linewidth]
    {figures/\figDir/box_7_naive_0001.png}%
    \adjincludegraphics[trim={{0.4\width} {0.0\height} {0.4\width} {0.0\height}}, clip, width=0.333\linewidth]
    {figures/\figDir/box_7_naive_0100.png}%
    \adjincludegraphics[trim={{0.4\width} {0.0\height} {0.4\width} {0.0\height}}, clip, width=0.333\linewidth]
    {figures/\figDir/box_7_naive_0300.png}%
    \caption{Naive initialization}
\end{subfigure}%
\begin{subfigure}[b]{0.333\linewidth}
    \adjincludegraphics[trim={{0.4\width} {0.0\height} {0.4\width} {0.0\height}}, clip, width=0.333\linewidth]
    {figures/\figDir/box_7_no_box_0001.png}%
    \adjincludegraphics[trim={{0.4\width} {0.0\height} {0.4\width} {0.0\height}}, clip, width=0.333\linewidth]
    {figures/\figDir/box_7_no_box_0100.png}%
    \adjincludegraphics[trim={{0.4\width} {0.0\height} {0.4\width} {0.0\height}}, clip, width=0.333\linewidth]
    {figures/\figDir/box_7_no_box_0300.png}%
    \caption{No box constraints}
\end{subfigure}%
\begin{subfigure}[b]{0.333\linewidth}
    \adjincludegraphics[trim={{0.4\width} {0.0\height} {0.4\width} {0.0\height}}, clip, width=0.333\linewidth]
    {figures/\figDir/box_7_with_box_0001.png}%
    \adjincludegraphics[trim={{0.4\width} {0.0\height} {0.4\width} {0.0\height}}, clip, width=0.333\linewidth]
    {figures/\figDir/box_7_with_box_0100.png}%
    \adjincludegraphics[trim={{0.4\width} {0.0\height} {0.4\width} {0.0\height}}, clip, width=0.333\linewidth]
    {figures/\figDir/box_7_with_box_0300.png}%
    \caption{With box constraints}
\end{subfigure}
\begin{subfigure}[b]{0.333\linewidth}
    \adjincludegraphics[trim={{0.4\width} {0.0\height} {0.4\width} {0.0\height}}, clip, width=0.333\linewidth]
    {figures/\figDir/box_8_naive_0001.png}%
    \adjincludegraphics[trim={{0.4\width} {0.0\height} {0.4\width} {0.0\height}}, clip, width=0.333\linewidth]
    {figures/\figDir/box_8_naive_0100.png}%
    \adjincludegraphics[trim={{0.4\width} {0.0\height} {0.4\width} {0.0\height}}, clip, width=0.333\linewidth]
    {figures/\figDir/box_8_naive_0300.png}%
    \caption{Naive initialization}
\end{subfigure}%
\begin{subfigure}[b]{0.333\linewidth}
    \adjincludegraphics[trim={{0.4\width} {0.0\height} {0.4\width} {0.0\height}}, clip, width=0.333\linewidth]
    {figures/\figDir/box_8_no_box_0001.png}%
    \adjincludegraphics[trim={{0.4\width} {0.0\height} {0.4\width} {0.0\height}}, clip, width=0.333\linewidth]
    {figures/\figDir/box_8_no_box_0100.png}%
    \adjincludegraphics[trim={{0.4\width} {0.0\height} {0.4\width} {0.0\height}}, clip, width=0.333\linewidth]
    {figures/\figDir/box_8_no_box_0300.png}%
    \caption{No box constraints}
\end{subfigure}%
\begin{subfigure}[b]{0.333\linewidth}
    \adjincludegraphics[trim={{0.4\width} {0.0\height} {0.4\width} {0.0\height}}, clip, width=0.333\linewidth]
    {figures/\figDir/box_8_with_box_0001.png}%
    \adjincludegraphics[trim={{0.4\width} {0.0\height} {0.4\width} {0.0\height}}, clip, width=0.333\linewidth]
    {figures/\figDir/box_8_with_box_0100.png}%
    \adjincludegraphics[trim={{0.4\width} {0.0\height} {0.4\width} {0.0\height}}, clip, width=0.333\linewidth]
    {figures/\figDir/box_8_with_box_0300.png}%
    \caption{With box constraints}
\end{subfigure}%
\vspace{-4mm}
\caption{
Box constraint evaluation with a single hair-like strand with $\stretchCoef = 10^7$. (Top) $\bendCoef = \twistCoef = 2.5 \times 10^6$. (Bottom) $\bendCoef = \twistCoef = 2.5 \times 10^7$. Our rest shape optimization without box constraints can significantly change the rest shape parameters for softer materials, driving the strands to have unstable local minima initially. Imposed box constraints can restrict changes in the rest shape parameters to achieve more stable dynamics and states while preserving the original shapes better than the naive initialization.
\vspace{-3mm}
}
\label{fig:box_constraints}
\end{figure}

\vspace{-4mm}
\subsection{Box Constraint Evaluation}
\label{sec:box_constraint_evaluation}
\vspace{-2mm}
To evaluate the effectiveness of the box constraints, we experiment with the hair-like strand, as shown in Figure \ref{fig:box_constraints}. We use two different sets of stiffness parameters ($\bendCoef = \twistCoef = 2.5 \times 10^6$ and $\bendCoef = \twistCoef = 2.5 \times 10^7$) and employ $\Delta \curvature = 0.6$ and $\Delta \curvature = 1.0$ in \eqref{eq:b}, respectively, given the very soft materials. We compare the following schemes:
\begin{enumerate}
\item Naive initialization;
\item No box constraints: initialized with our rest shape optimization using no box constraints;
\item With box constraints.
\end{enumerate}

Without using the box constraints, the rest shape parameters can be modified significantly, leading to a relatively unstable static equilibrium and thus failing to retain the original strand shape. In particular, for softer materials, the rest shape change can be excessive, introducing stability issues in the forward simulation. By contrast, the use of box constraints can avoid introducing stability problems and achieve stable strand dynamics and states \mycolor{(although the strand needs to compromise static equilibrium)}, preserving the original shape better compared to the naive initialization. 


\begin{figure}[tb]
\captionsetup[subfigure]{aboveskip=0mm,belowskip=0mm,font=small}
\centering
\begin{subfigure}[b]{0.333\linewidth}
    \adjincludegraphics[trim={{0.4\width} {0.1\height} {0.35\width} {0.0\height}}, clip, width=0.333\linewidth]
    {figures/\figDir/cond_f2_0001.png}%
    \adjincludegraphics[trim={{0.4\width} {0.1\height} {0.35\width} {0.0\height}}, clip, width=0.333\linewidth]
    {figures/\figDir/cond_f2_0035.png}%
    \adjincludegraphics[trim={{0.4\width} {0.1\height} {0.35\width} {0.0\height}}, clip, width=0.333\linewidth]
    {figures/\figDir/cond_f2_0075.png}%
    \caption{Force norm}
\end{subfigure}%
\begin{subfigure}[b]{0.333\linewidth}
    \adjincludegraphics[trim={{0.4\width} {0.1\height} {0.35\width} {0.0\height}}, clip, width=0.333\linewidth]
    {figures/\figDir/cond_scale_0001.png}%
    \adjincludegraphics[trim={{0.4\width} {0.1\height} {0.35\width} {0.0\height}}, clip, width=0.333\linewidth]
    {figures/\figDir/cond_scale_0035.png}%
    \adjincludegraphics[trim={{0.4\width} {0.1\height} {0.35\width} {0.0\height}}, clip, width=0.333\linewidth]
    {figures/\figDir/cond_scale_0075.png}%
    \caption{Equilibration}
\end{subfigure}%
\begin{subfigure}[b]{0.333\linewidth}
    \adjincludegraphics[trim={{0.4\width} {0.1\height} {0.35\width} {0.0\height}}, clip, width=0.333\linewidth]
    {figures/\figDir/cond_ours_0001.png}%
    \adjincludegraphics[trim={{0.4\width} {0.1\height} {0.35\width} {0.0\height}}, clip, width=0.333\linewidth]
    {figures/\figDir/cond_ours_0035.png}%
    \adjincludegraphics[trim={{0.4\width} {0.1\height} {0.35\width} {0.0\height}}, clip, width=0.333\linewidth]
    {figures/\figDir/cond_ours_0075.png}%
    \caption{Kinetic energy norm}
\end{subfigure}
\vspace{-4mm}
\caption{
\mycolor{
Experiment with a hair-like strand. Optimization with the force norm and equilibration approach fails to achieve static equilibrium whereas our method succeeds.
}
\vspace{-6mm}
}
\label{fig:conditioning}
\end{figure}

\begin{table}[tb]
\centering
\caption{
\mycolor{
Results at solver convergence for the experiment in Figure \ref{fig:conditioning}. $\mathrm{min}(\Amat)$ and $\mathrm{max}(\Amat)$ denote the minimum and maximum magnitude of non-zeros in the system matrix $\Amat$, respectively, $\sigma$ their ratio ($\mathrm{max}(\Amat)/\mathrm{min}(\Amat)$), $\norm{\generalizedForce}_2$ and $\norm{\generalizedForce}_{\generalizedMass^{-1}}$ the norm of corresponding objectives.
\vspace{-3mm}
}
}
\scalebox{0.65}{
\begin{tabular}{c|rrrrrrrrrr}\hline
Schemes & $\mathrm{min}(\Amat)$ & $\mathrm{max}(\Amat)$ & $\sigma$ & $\norm{\generalizedForce}_2$ & $\norm{\generalizedForce}_{\generalizedMass^{-1}}$\\
\hline \hline
Force norm                       & $6.0\times 10^{-9}$ & $8.4\times 10^{8}$ & $1.4\times 10^{17}$ & $3.1\times 10^{-4}$ & $1.2\times 10^{1}$ \\
Equilibration           & $1.9\times 10^{-14}$ & $1.0\times 10^{0}$ & $5.4\times 10^{13}$ & $3.1\times 10^{-4}$ & $1.2\times 10^{1}$ \\
Kinetic energy norm & $1.4\times 10^{-2}$ & $2.7\times 10^{13}$ & $1.9\times 10^{15}$ & $4.3\times 10^{-8}$ & $6.3\times 10^{-6}$ \\
\hline
\end{tabular}}
\label{tab:conditioning}
\vspace{-3mm}
\end{table}

\vspace{-4mm}
\subsection{\mycolor{Evaluation of Norm Formulations}}
\label{sec:conditioning}
\vspace{-2mm}
\mycolor{
To demonstrate the effectiveness of our norm formulation, we experiment with the hair-like strand, as shown in Figure \ref{fig:conditioning}. We compare the following schemes:
\begin{enumerate}
\item Force norm: force-norm formulation $\norm{\generalizedForce}_2^2$ \cite{Twigg:2011:OSS:2019406.2019437} which forms the Hessian via $\Jacobian^T \Jacobian$;
\item Equilibration: force-norm formulation \cite{Twigg:2011:OSS:2019406.2019437} with symmetric equilibration \cite{Bunch1971} to improve numerical conditioning of the system via $\Dmat \Jacobian^T \Jacobian \Dmat$ ($\Dmat$: diagonal scaling matrix);
\item Kinetic energy norm: our kinetic-energy-based norm $\norm{\generalizedForce}_{\generalizedMass^{-1}}^2$ which forms the Hessian via $\Jacobian^T \generalizedMass^{-1} \Jacobian$.
\end{enumerate}
We ignore the constant scaling of objectives for readability. Table \ref{tab:conditioning} summarizes the result of the experiment. Optimization with the force norm fails to achieve static equilibrium because the system has entries significantly different in magnitude and thus neglects some force components. While equilibrating the matrix improved the numerical conditioning of the system with smaller variations in the matrix entries (i.e., smaller $\sigma$), it did not help to further reduce the objective values within the limited numerical precision (giving a result almost identical to the one with the force norm), because it neglects the energy contributions from each force component. By contrast, our kinetic-energy-based objective properly considers the energy contributions in the system and sufficiently reduces both the objectives $\norm{\generalizedForce}_2^2$ and $\norm{\generalizedForce}_{\generalizedMass^{-1}}^2$, thus achieving static equilibrium.
}


\begin{figure}[tb]
\captionsetup[subfigure]{aboveskip=0mm,belowskip=0mm,font=small}
\centering
\begin{subfigure}[b]{1.0\linewidth}
    \adjincludegraphics[trim={{0.1\width} {0.0\height} {0.1\width} {0.15\height}}, clip, width=0.25\linewidth]
    {figures/\figDir/08_coil_wo_0001.png}%
    \adjincludegraphics[trim={{0.1\width} {0.0\height} {0.1\width} {0.15\height}}, clip, width=0.25\linewidth]
    {figures/\figDir/08_coil_wo_0023.png}%
    \adjincludegraphics[trim={{0.1\width} {0.0\height} {0.1\width} {0.15\height}}, clip, width=0.25\linewidth]
    {figures/\figDir/08_coil_wo_0534.png}%
    \adjincludegraphics[trim={{0.1\width} {0.0\height} {0.1\width} {0.15\height}}, clip, width=0.25\linewidth]
    {figures/\figDir/08_coil_wo_0900.png}%
    \caption{Naive initialization}
\end{subfigure}
\begin{subfigure}[b]{1.0\linewidth}
    \adjincludegraphics[trim={{0.1\width} {0.0\height} {0.1\width} {0.15\height}}, clip, width=0.25\linewidth]
    {figures/\figDir/08_coil_w_0001.png}%
    \adjincludegraphics[trim={{0.1\width} {0.0\height} {0.1\width} {0.15\height}}, clip, width=0.25\linewidth]
    {figures/\figDir/08_coil_w_0023.png}%
    \adjincludegraphics[trim={{0.1\width} {0.0\height} {0.1\width} {0.15\height}}, clip, width=0.25\linewidth]
    {figures/\figDir/08_coil_w_0534.png}%
    \adjincludegraphics[trim={{0.1\width} {0.0\height} {0.1\width} {0.15\height}}, clip, width=0.25\linewidth]
    {figures/\figDir/08_coil_w_0900.png}%
    \caption{With rest shape optimization}
\end{subfigure}%
\vspace{-4mm}
\caption{
\mycolor{Evaluation using a coiled hair-like strand, with its root vertex perturbed. The naive initialization fails to achieve static equilibrium and suffers from the stability issue, whereas the rest shape optimization enables sag-free and stable simulation.
}
\vspace{-7mm}
}
\label{fig:coil}
\end{figure}

\vspace{-4mm}
\subsection{\mycolor{Evaluation with Coiled Hair-Like Strand}}
\vspace{-2mm}
\mycolor{
To demonstrate that our method can handle curly strands, we experiment with a coiled hair-like strand, as shown in Figure \ref{fig:coil}, using $\stretchCoef = \bendCoef = \twistCoef = 10^{9}$. By simultaneously optimizing the rest length, rest curvature, and rest twist, our method achieves static equilibrium. Additionally, when the root vertices are perturbed, the optimized rest shape prevents too much strand deformation and thus stabilizes the simulation in this example, whereas the strand with the naive initialization exhibits unnatural oscillations.
}


\begin{figure*}[tb]
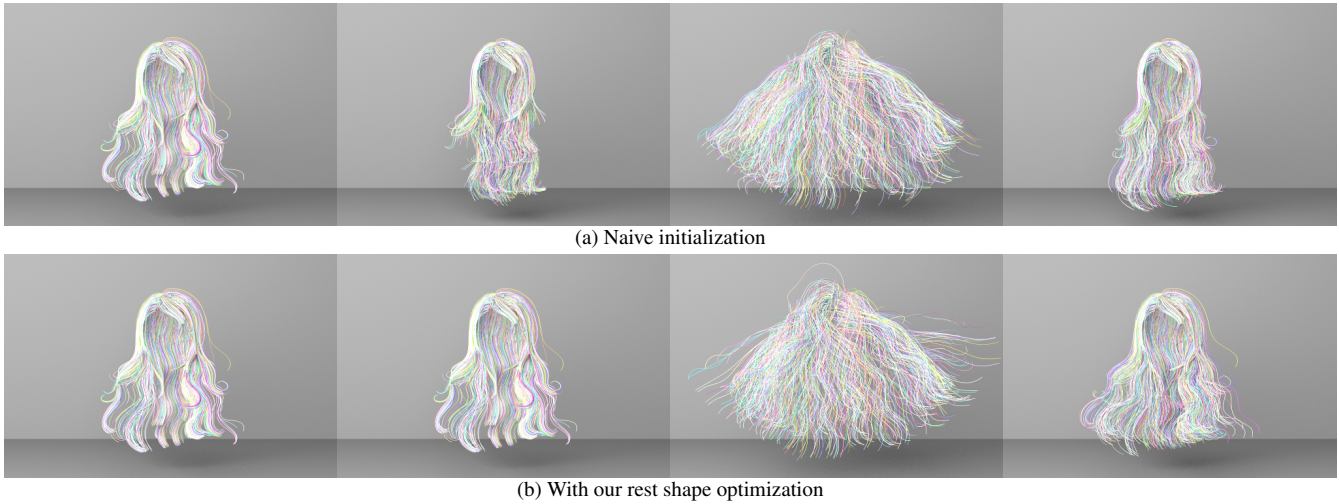

\captionsetup[subfigure]{aboveskip=0mm,belowskip=0mm,font=small}
\centering
\begin{subfigure}[b]{1.0\linewidth}
    \adjincludegraphics[trim={{0.1\width} {0.05\height} {0.1\width} {0.0\height}}, clip, width=0.25\linewidth]
    {figures/\figDir/hairs_without_0001.png}%
    \adjincludegraphics[trim={{0.1\width} {0.05\height} {0.1\width} {0.0\height}}, clip, width=0.25\linewidth]
    {figures/\figDir/hairs_without_0030.png}%
    \adjincludegraphics[trim={{0.1\width} {0.05\height} {0.1\width} {0.0\height}}, clip, width=0.25\linewidth]
    {figures/\figDir/hairs_without_0115.png}%
    \adjincludegraphics[trim={{0.1\width} {0.05\height} {0.1\width} {0.0\height}}, clip, width=0.25\linewidth]
    {figures/\figDir/hairs_without_0800.png}%
    \caption{Naive initialization}
\end{subfigure}
\begin{subfigure}[b]{1.0\linewidth}
    \adjincludegraphics[trim={{0.1\width} {0.05\height} {0.1\width} {0.0\height}}, clip, width=0.25\linewidth]
    {figures/\figDir/hairs_with_0001.png}%
    \adjincludegraphics[trim={{0.1\width} {0.05\height} {0.1\width} {0.0\height}}, clip, width=0.25\linewidth]
    {figures/\figDir/hairs_with_0030.png}%
    \adjincludegraphics[trim={{0.1\width} {0.05\height} {0.1\width} {0.0\height}}, clip, width=0.25\linewidth]
    {figures/\figDir/hairs_with_0115.png}%
    \adjincludegraphics[trim={{0.1\width} {0.05\height} {0.1\width} {0.0\height}}, clip, width=0.25\linewidth]
    {figures/\figDir/hairs_with_0800.png}%
    \caption{With our rest shape optimization}
\end{subfigure}%
\vspace{-4mm}
\caption{
Hair simulation with \mycolor{hair assets}. Hair strands sag with the naive initialization while our method enables hair strands to retain their original shapes, exhibit natural motions due to the prescribed motion of the root vertices, and then return toward the original shape.
\vspace{-7mm}
}
\label{fig:hairs}
\end{figure*}

\begin{figure}[tb]
\captionsetup[subfigure]{aboveskip=0mm,belowskip=0mm,font=small}
\centering
\begin{subfigure}[b]{0.333\linewidth}
    \adjincludegraphics[trim={{0.25\width} {0.0\height} {0.25\width} {0.1\height}}, clip, width=\linewidth]
    {figures/\figDir/hair_thin_initial_0001.png}%
    \caption{Initial condition}
\end{subfigure}%
\begin{subfigure}[b]{0.333\linewidth}
    \adjincludegraphics[trim={{0.25\width} {0.0\height} {0.25\width} {0.1\height}}, clip, width=\linewidth]
    {figures/\figDir/hair_thin_without_0800.png}%
    \caption{Naive initialization}
\end{subfigure}%
\begin{subfigure}[b]{0.333\linewidth}
    \adjincludegraphics[trim={{0.25\width} {0.0\height} {0.25\width} {0.1\height}}, clip, width=\linewidth]
    {figures/\figDir/hair_thin_with_0800.png}%
    \caption{With optimization}
\end{subfigure}%
\vspace{-4mm}
\caption{
\mycolor{Experiment with thinner strands. With naive initialization, strands sag significantly. With rest shape optimization, strands still sag (though sagging is mitigated) and exhibit unnatural lifting.}
\vspace{-9mm}
}
\label{fig:failure_case}
\end{figure}

\vspace{-4mm}
\subsection{Evaluation with Complex Strand Geometry}
\vspace{-2mm}
To evaluate the effectiveness of our method in more complex scenarios, we experiment with hair strand data released publicly by Hu et al. \cite{Hu2015hair}. We use $\stretchCoef = 3.0 \times 10^{8} , \bendCoef = 3.0 \times 10^{8}$, and $\twistCoef = 3.0 \times 10^{8}$. Figure \ref{fig:hairs} compares our method with the naive initialization. While the hair strands sag due to the gravity with the naive initialization, the strands with our method retain the original hair style. When the root vertices are rotated in a prescribed way, both approaches generate natural and comparable hair motions. After the root vertices are stopped at the same position as at the start, hair strands with our approach retain the original hair shape, unlike those with the naive initialization. Figure \ref{fig:teaser} shows hair simulations using our method with another complex hair style.

\mycolor{In Figure \ref{fig:failure_case}, we use the same setting as Figure \ref{fig:hairs}, but with thinner strands (strand radius $\radius = 10^{-4}$). In this case, rest shape optimization fails to retain the original hair shape (while the sagging is mitigated compared to naive initialization) because the strands cannot exert sufficiently strong forces to support themselves. Additionally, significant rest shape changes can lead to unstable strand configurations, causing unnatural hair lifting.}



\vspace{-4mm}
\section{Conclusions and Future Work}
\vspace{-2mm}
We have proposed our rest shape optimizer to achieve sag-free DER simulation and evaluated its efficacy in various examples. In the following, we discuss tradeoffs inherent to our approach and promising research directions for future work.


\vspace{-4mm}
\subsection{Static Equilibrium at Local Maxima and Minima}
\vspace{-2mm}
Our formulation based on the kinetic energy is akin to the force-norm minimization \cite{Twigg:2011:OSS:2019406.2019437} (which is also related to the nonlinear force solves with inverse dynamics \cite{Hadap2006strands,Featherstone2016}, ANM \cite{Chen2014asymptotic,Jia2021SANM}, and global-local approaches \cite{Hsu2022sag,Hsu2023sag}) and is designed to find a static equilibrium where forces are zero. While the achieved static equilibrium is at a local minimum for the rest shape optimization \eqref{eq:rest_shape_optimization_unconstrained}, it can be at a local maximum for the optimization on forward simulation \eqref{eq:forward_sim_optimization} \cite{Derouet-Jourdan2010,Hsu2022sag}. As such, perturbations on the generalized positions can relatively easily break the static equilibrium; however, this case is rare for strands with one-sided clamping since such strands typically do not experience compression and settle at a local minimum for \eqref{eq:forward_sim_optimization} via the rest shape optimization \eqref{eq:rest_shape_optimization_unconstrained}. If one needs to guarantee a static equilibrium at a local minimum with respect to both generalized positions and rest shape parameters, it seems promising to use local material stiffening \cite{Derouet-Jourdan2010,Hsu2022sag} and the adjoint-based approaches \cite{Perez2015rod}.


\vspace{-4mm}
\subsection{Material Stiffness Parameter Optimization}
\vspace{-2mm}
While our rest shape optimizer was able to find rest shape parameters that achieve the desired static equilibrium in various settings with reasonable material stiffness parameters, in general, it is not guaranteed that such parameters exist, e.g., when the given system is overdetermined or such parameters are outside of the box constraints \mycolor{(see Figures \ref{fig:horizontal} and \ref{fig:box_constraints})}. In addition, when the rest shape is significantly modified to achieve the static equilibrium (even under the box constraints, which are designed to avoid introducing stability problems), the strand shape may not return to the original configuration after the strand moves dynamically because it can get stuck in different local extrema \eqref{eq:forward_sim_optimization}. In these cases, increasing the material stiffness \mycolor{(or equivalently increasing the strand radius)} can reduce the required rest shape changes and can be helpful to establish a stabler static equilibrium (albeit compromising user-desired elastic dynamics)\cite{Derouet-Jourdan2010,Twigg:2011:OSS:2019406.2019437,Hsu2022sag,Hsu2023sag}.


\vspace{-4mm}
\subsection{Optimization Solvers}
\vspace{-2mm}
We designed our optimization solvers by seeking a balance between performance and simplicity, but it is possible that other alternatives could potentially perform better. Although the penalty method was quite satisfactory in our examples (since small violations of box constraints or clamping were permissible), other constrained optimization solvers (e.g., augmented Lagrangian method (ALM) \cite{NoceWrig06,Takahashi2021}, interior point method (IPM) \cite{Takahashi2024qp}, and active-set method \cite{Dostal2005,Takahashi2023}) could be employed to more accurately satisfy the box constraints. Other nonlinear solvers (e.g., nonlinear conjugate gradient, quasi-Newton, and LMA \cite{NoceWrig06}) are also available, and it would be worth comparing their performance with Gauss-Newton. While we arranged the rest shape parameters in their current order (rest length, rest curvature, rest twist) so that the rest twist can easily be added and removed from the rest shape optimization, by interleaving these parameters (as we did for the generalized positions),  the inner linear systems for Newton-type optimizers would possess a banded structure that enables more efficient direct solves without reordering of the system(s). 


\vspace{-4mm}
\subsection{Toward More General Inverse Problems}
\vspace{-2mm}
In our framework, for simplicity and efficiency, we assumed a strand with constant radius, density, and material stiffness. It would be worthwhile removing this assumption to support more general anisotropic and inhomogeneous strands \cite{Hafner2021curves,Hafner2023rod}. Moreover, our framework should be able to extend to support two-dimensional shells and three-dimensional volumetric structures, along with more rigorous analysis for the box constraints imposed on the rest shape parameters. In addition, although we also assumed no contacts among strands or other objects, this approach can lead to larger rest shape changes than necessary in cases where the elastic strands could be supported by each other or other objects when the simulation starts \cite{Derouet-Jourdan:2013:IDH:2508363.2508398,Ly2018shell,Hsu2022sag,Hsu2023sag}. Thus, a promising extension would be to take frictional contacts into account during the rest shape optimization.


\vspace{-4mm}
\section*{Acknowledgements}
\vspace{-2mm}
We thank the anonymous reviewers for their valuable suggestions and comments, which significantly improved this work. We are also grateful to Bo Yang, Qingyuan Zheng, Rundong Wu, and Yu Ju Chen for their discussions during the early stages of this research. Additionally, we acknowledge the authors of \cite{Hu2015hair} for publicly releasing the hair dataset. This work was supported in part by the Natural Sciences and Engineering Research Council of Canada (Grant RGPIN-2021-02524).

\vspace{-4mm}
\bibliographystyle{eg-alpha-doi} 
\bibliography{source/reference}       

\end{document}